%% file: apssamp.tex
\documentclass[%
 reprint,
 amsmath,amssymb,
 aps,
 prd,
floatfix,
]{revtex4-2}

\usepackage{graphicx}
\usepackage{dcolumn}
\usepackage{bm}
\usepackage{hyperref}

\usepackage{xspace} 
\usepackage{booktabs}

\input{symbols-def}

\usepackage{orcidlink}

\begin{document}

\preprint{APS/???}

\title{Constraints on new physics from decays of polarized \texorpdfstring{\Lb}{Lb} baryons at the FCC-ee}
\thanks{Contact author: anbeck@mit.edu}
\author{Anja Beck$^1$\orcidlink{0000-0003-4872-1213}}
\author{Mero Elmarassy$^1$}
\author{Asher Sabbagh$^1$}
\author{Michal Kreps$^2$\orcidlink{0000-0002-6133-486X}}
\author{Eluned Smith$^1$\orcidlink{0000-0002-9740-0574}}
\affiliation{\it\small$^1$Department of Physics and Laboratory for Nuclear Science, MIT, Cambridge, 02139, MA, USA \\
\it\small$^2$Department of Physics, University of Warwick, Coventry, CV4 7AL, West Midlands, United Kingdom}

\date{\today}

\begin{abstract}
The $Z^0$ bosons produced in electron-positron collisions at the potential Future Circular Collider (FCC-ee) provide unique opportunities for flavor physics.
The nonzero polarization of \Lb baryons produced in 
$Z^0$ decays enables access to a much larger set of observables than at the LHC, where the \Lb baryons are produced unpolarized.
This paper presents a toy angular analysis of \LbToppimm decays using simulation samples of collisions at the FCC-ee reconstructed using the IDEA detector concept and assuming a dataset of $6\times 10^{12}$ $Z^0$ bosons .
While the statistical sensitivity achieved for individual angular observables is not expected to significantly exceed that from the LHCb Upgrade~II experiment, the addition of the polarized observables leads to a significant improvement of the knowledge on the Wilson coefficients $C_{9^{(\prime)}}$ and $C_{10^{(\prime)}}$.
\end{abstract}

\maketitle

\section{Introduction}
Electroweak penguin decays of heavy hadrons provide a powerful window into the structure of the Standard Model (SM) and offer sensitivity to potential new physics effects that may appear at energy scales beyond the direct reach of current or future colliders.
There have been long-standing tensions observed between measurement and prediction in \bsll decays, in particular \BdbToKstmm~\cite{CMS:2024atz,Belle:2016fev,ATLAS:2018gqc,LHCb-PAPER-2020-002}.
Other hadronic systems have been studied also and often exhibit similar tensions with the SM~\cite{LHCb-PAPER-2013-017,LHCb-PAPER-2015-023,LHCb-PAPER-2021-014, LHCb-PAPER-2014-006,LHCb-PAPER-2020-041,LHCb-PAPER-2023-033,LHCb-PAPER-2024-011,CMS:2024syx, BaBar:2012mrf, Belle:2019xld,CMS:2024syx,LHCb-PAPER-2014-024,LHCb-PAPER-2017-013,LHCb-PAPER-2019-040,LHCb-PAPER-2021-004,LHCb-PAPER-2021-038,LHCb-PAPER-2022-045,LHCb-PAPER-2022-046}.
Due to their nonzero spin, decays of baryons, such as \LbToppimm, provide very rich angular structures.
Additional observables become accessible in the case of nonzero polarization and enhance the sensitivity to different helicity structures of the underlying effective operators, allowing for a more nuanced disentanglement of the Wilson coefficients, see, e.g., Ref.~\cite{Detmold:2012vy}.

The combination of experiments at the Large Electron Positron collider (LEP) observed a significant longitudinal \Lb polarization of $P_\parallel=-0.45$ in collisions of unpolarized electron and positron beams at a center-of-mass energy of $\sqrt{s}=m_{Z^0}c^2=91$\gev~\cite{ALEPH:2001ccp}.
The proposed Future Circular Collider (FCC-ee)~\cite{benedikt_2025_15487832}, operating as a high-luminosity $Z^0$ factory, provides a unique opportunity to perform detailed studies of these polarized \Lb baryons.
This includes in particular the measurement of angular observables, which are inaccessible at hadron colliders due to the strong production of \bbbar pairs leading to unpolarized \Lb baryons.

This publication estimates the sensitivity on angular observables in \LbToppimm decays, using a maximum likelihood fit, to be expected by the end of the FCC-ee run period.
The increase in sensitivity to the Wilson coefficients using the full set of observables over the unpolarized set is also determined.
A future study may focus on decays to other leptons, which provide great opportunity at the FCC-ee given the simpler environment compared to the LHC.

\section{Angular distribution}
The differential decay rate of \LbToppimm is~\cite{Blake:2017une}
\begin{equation}
\begin{split}
&\frac{32\pi^{2}}{3} \frac{\deriv^{6}\Gamma}{\deriv\qsq\,\deriv\vec{\Omega}} = \\
&\quad \left(K_1\sin^2\tm+K_2\cos^2\tm+K_3\cos\tm\right) \\
&+ \left(K_4\sin^2\tm+K_5\cos^2\tm+K_6\cos\tm\right)\cos\th \\
&+ \left(K_7\sin\tm\cos\tm+K_8\sin\tm\right)\sin\th\cos\left(\phih+\phim\right) \\
&+ \left(K_9\sin\tm\cos\tm+K_{10}\sin\tm\right)\sin\th\sin\left(\phih+\phim\right) \\
&+ \left(K_{11}\sin^2\tm+K_{12}\cos^2\tm+K_{13}\cos\tm\right)\cos\tp \\
&+ \left( K_{14}\sin^2\tm+K_{15}\cos^2\tm+K_{16}\cos\tm\right)\cos\th \cos\tp \\
&+ \left(K_{17}\sin\tm\cos\tm+K_{18}\sin\tm\right)\sin\th\cos\left(\phih+\phim\right)\cos\tp  \\
&+ \left(K_{19}\sin\tm\cos\tm+K_{20}\sin\tm\right)\sin\th\sin\left(\phih+\phim\right) \cos\tp \\
&+ \left(K_{21}\cos\tm\sin\tm+K_{22}\sin\tm\right)\sin\phim \sin\tp \\
&+ \left(K_{23}\cos\tm\sin\tm+K_{24}\sin\tm\right)\cos\phim  \sin\tp \\
&+ \left(K_{25}\cos\tm\sin\tm+K_{26}\sin\tm\right)\sin\phim\cos\th  \sin\tp \\
&+ \left(K_{27}\cos\tm\sin\tm+K_{28}\sin\tm\right)\cos\phim\cos\th  \sin\tp  \\
&+ \left(K_{29}\cos^2\tm+K_{30}\sin^2\tm\right)\sin\th\sin\phih  \sin\tp \\
&+ \left(K_{31}\cos^2\tm+K_{32}\sin^2\tm\right)\sin\th\cos\phih  \sin\tp \\
&+ \left(K_{33}\sin^2\tm \right) \sin\th\cos\left(2\phim+\phih\right) \sin\tp  \\
&+ \left(K_{34}\sin^2\tm \right) \sin\th\sin\left(2\phim+\phih\right)  \sin\tp ~,
\end{split}
\label{eq:decayrate}
\end{equation}
where the dependence on the dimuon invariant mass squared, \qsq, is absorbed into the angular coefficients $K_i$ and the five-dimensional angular phase space is given by \mbox{$\vec{\Omega}=(\cos\tp,\cos\tm,\cos\th,\phim,\phih)$}.
The angle between the \Lb momentum in the lab frame and the \Lz momentum in the \Lb rest frame defines the longitudinal polarization angle \tp.
The muon helicity angle, \tm, corresponds to the angle between the momentum of the muon with the same charge as the proton in the dimuon rest frame and the dimuon momentum in the \Lb rest frame.
Similarly, the hadron helicity angle, \th, is the angle between the proton momentum in the dihadron rest frame and the dihadron momentum in the \Lb rest frame.
The azimuthal angles $\phim$ and $\phih$ are the angles between the muon and hadron decay planes and the \Lb momentum in the lab frame.
The first ten coefficients are independent of the \Lb polarization.
In this case, the angular space collapses to three dimensions as the polarization angle disappears from the decay rate and the azimuthal angles only appear together as $\phi=\phim+\phih$.
For an illustration of the angular definitions, please refer to Ref.~\cite{Blake:2017une} or similar publications.
Note that sign conventions may differ depending on which particles are chosen as reference.

\section{Simulation and detector response}
Simulation samples of collisions of unpolarized electron and positron beams at the $Z^0$ pole decaying to \qqbar pairs are generated with Pythia~\cite{Sjostrand:2014zea}.
The \LbToppimm signal as well as the dominant background, \BdTopipimm, are decayed according to phase space availability using EvtGen~\cite{Lange:2001uf}.
Realistic distributions resembling the Standard Model are achieved using weights.
The \LbToppimm model for the reweighting employs the Wilson coefficients from EvtGen~\cite{Buras:1993xp,Buras:1994dj} and lattice QCD calculations for the $\Lb\to\Lz$ hadronic form factors~\cite{Detmold:2012vy}.
The longitudinal \Lb polarization has been set to $P_\parallel=-0.45$ as measured at LEP~\cite{ALEPH:2001ccp} in all studies shown later.
Appendix~\ref{app:distributions} (Fig.~\ref{fig:distributions}) shows the resulting one-dimensional \Lb decay distributions before and after applying the model weights to the simulation.
Different values for the polarization, including nonzero transverse polarization are sometimes used for cross-checks.
The \BdTopipimm background samples are reweighted using the angular decay rate~\cite{Bobeth:2011nj,LHCb-PAPER-2014-007}
\begin{align}
    \frac{\deriv\Gamma}{\deriv\cos\tm} = \frac{3}{4}\left(1-\cos^2\tm\right) \ ,
\end{align}
which only depends on the muon helicity angle $\tm$.
The dependence of the decay rate on the dimuon invariant mass is kept as phase space.

This study considers the Innovative Detector for Electron-positron Accelerators (IDEA) concept~\cite{IDEAStudyGroup:2025gbt}.
The IDEA detector features a silicon-pixel vertex detector, a large-volume extremely light short-drift wire chamber surrounded by a layer of silicon microstrip detectors, a finely segmented crystal electromagnetic calorimeter, a thin low-mass superconducting solenoid coil, a dual-readout fiber  calorimeter, and muon chambers within the magnet return yoke.
The detector response is emulated using the DELPHES fast simulation framework~\cite{deFavereau:2013fsa}.

\section{Selection and backgrounds}
The unique decay topology of the \LbToppimm decay, caused by the long lifetimes of the \Lb and \Lz baryons, provides useful features for an efficient signal selection.
Decay candidates are reconstructed by first finding all vertices that are not primary and have either exactly two muon tracks or exactly two hadron tracks, in both cases of opposite charge.
The muon identification as well as the ability to classify a track as a charged hadron is assumed to be perfect.
A real detector will have a small nonzero misidentification, rate which is however no concern as there are no background species with branching fractions large enough to pollute the sample in meaningful ways.
The invariant mass of the dihadron pair is calculated for the $p\pi$ and $\pi p$ hypotheses and particle masses are assigned based on which combination results in an invariant mass closer to the true \Lz mass.

Bunch crossings at the FCC-ee are projected to produce exactly one $Z^0$ boson which decays with a probability of around 15\% into a \bbbar pair~\cite{benedikt_2025_15487832,PDG2024}.
The probability of producing an additional \bbbar pair in the hadronization process is very low such that events with more than one $b$ quark and one anti-$b$ quark are neglected in this study~\cite{OPAL:2000wtx}.
As a consequence in every event, there are a maximum of one \Lb and one \Lbbar baryon that fly back to back in the lab frame.
The combination of the dihadron and dimuon pair with the smallest angle between the momentum of the \Lz and the \Lb candidates (or \Lbar and \Lbbar) is chosen as the \Lb (or \Lbbar) candidate in each event.
Note however that due to the small branching fraction of $\mathcal{B(\LbToppimm)}=6.9\cdot 10^{-7}$~\cite{PDG2024}, the chances of getting both a \Lb and \Lbbar signal decay in one event are vanishingly small.

Backgrounds arising from wrongly selected combinations of unrelated dimuon and dihadron pairs can be suppressed by imposing that the flight distance of the \Lb (calculated from the distance between the primary vertex and the dimuon vertex) is at least 100 times as large as the impact parameter of the \Lb candidate with respect to the primary vertex.
Requiring that the \Lz flies at least 10\mm allows one to neglect backgrounds from other rare decays involving strongly decaying resonances such as \BdbToKstmm.
Considering only a narrow window around the \Lb (100\mevcc) and \Lz (20\mevcc) masses additionally suppresses different sources of background.
See Appendix~\ref{app:resolution} (Fig.~\ref{fig:width}) for the resolution of the invariant masses.

The analysis is performed in two bins of dimuon invariant mass where bin 1 covers the region between the $\phi$ and \jpsi resonances and bin 2 covers the region above the \psitwos resonance up to just below the edge of the phase space.
The region at very low dimuon invariant mass is neglected for two reasons.
First, the model used to reweight the signal simulation has a strong photon pole at the dimuon threshold, see Appendix~\ref{app:distributions} (Fig.~\ref{fig:distributions}), which is recreated by assigning very large weights to the small number of data points present in this region of the phase space.
These weights are one to two orders of magnitude larger than the weights in other regions potentially leading to instabilities in the fit.
Second, the $Z^0\to\qqbar$ background samples indicate a complex spectrum of low-mass dimuon resonances and combinatorial background in this region which is discussed below.
Given that the signal distribution in the low-mass region is dominated by the contribution of $\mathcal{C}_7$ which is very well known today, see, e.g., Refs.~\cite{LHCb-PAPER-2020-020,LHCb-PAPER-2024-030}, it is most economical to focus on the regions denominated bins 1 and 2 in this study.
The second bin is restricted to just below the kinematically allowed limit of dimuon invariant mass in order to obtain a useful efficiency model from the sparsely populated far edge of the phase space.
Table~\ref{tab:selections} provides a summary of all employed selections.

Rare \BdTopipimm decays pose the potentially most dangerous background given their similar topology to the signal.
Due to the roughly ten-times larger hadronization fraction of \Bd mesons compared to \Lb baryons, the \KS background contributes at a level of 11\% of the signal yield without applying any particle identification criteria.
Vetoing the \KS mass region cannot sufficiently suppress this background and leads to inconvenient localized efficiency dips.
As a consequence, no \KS veto is applied.
The impact of this background on the sensitivity of the angular observables is assessed for different levels of $p-\pi$ separation power as part of the systematic uncertainty.

\onecolumngrid
\begin{center}
\begin{table}
    \centering
    \caption{Selections suppressing background decays and definition of the analysis regions.}
    \label{tab:selections}
    \begin{tabular}{p{0.45\textwidth}|c}
         \toprule
         Purpose & Selection \\
         \midrule
         Reject wrong topologies
         & \begin{tabular}{@{}c@{}}
            $100\cdot\text{IP}(\Lb)<\text{FD}(\Lb)$ \\
            $10\mm<\text{FD}(\Lz)$
        \end{tabular} \\
        \midrule
        Suppress backgrounds far away from the signal region & \begin{tabular}{@{}c@{}}
            $|m(p\pi\mu\mu)-m_\Lb|<100\mevcc$ \\
            $|m(p\pi)-m_\Lz|<20\mevcc$
        \end{tabular} \\
        \midrule
        Analysis bin 1 & $m(\mu\mu)\in[m_\phi+60,m_\jpsi-60]\mevcc$ \\
        Analysis bin 2 & $m(\mu\mu)\in[m_\psitwos+60,m_\Lb-m_\Lz-20]\mevcc$ \\
        \bottomrule
    \end{tabular}
\end{table}
\end{center}
\twocolumngrid

Table~\ref{tab:contaminations} lists the potential sources of background investigated for this study compared to the signal.
The values for the \bquark hadron production and branching fractions are based on Refs.~\cite{HFLAV:2019otj,PDG2024}.
Note that for \bquark baryons in $Z^0$ decays only a combined production fraction of $f_{b\text{ baryon}}=0.085$ has been measured.
Assuming that the \bquark baryon spectrum is similar at the $Z^0$ pole compared to hadron colliders~\cite{Jiang:2018iqa}, our estimate for the \Lb baryon production fraction is $f_\Lb=0.075$.
The label \textit{Background with \LbToppimm} refers to backgrounds that arise within a signal event, for example when the dihadron pair stemming from a \Lb was matched with an unrelated dimuon pair.
The second column shows the expected number of events for each category assuming $6\cdot10^{12}$ $Z^0$ bosons by the end of the FCC-ee run period~\cite{FCC:2025uan,PDG2024}.
The following columns list the efficiency of the different selection steps where \textit{Acceptance} includes the efficiency due to the detector geometry as well as the reconstruction procedure, \textit{Topology} includes the acceptance as well as the topological criteria, and \textit{Total} includes all requirements listed in Table~\ref{tab:selections} on top of the acceptance.
Note that the efficiencies are calculated with respect to the number of generated events while the reconstruction allows for up to one \LbToppimm and one $\Lbbar\to\Lbar(\to \overline{\proton}\pip)\mumu$ candidate per event which can result in efficiencies larger than one.
The rightmost column indicates the expected number of candidates in the signal region of the final dataset.
In case there are zero candidates left, the efficiency and number of candidates are given as an upper limit at 68\% confidence level.
The efficiencies and estimated number of selected candidates stem from samples generated according to phase space availability for \LbToppimm and \BdTopipimm.
The projected yield of \LbToppimm candidates exceeds the expected number of candidates collected by the LHCb experiment by the end of Upgrade~I but does not supersede the yield including LHCb Upgrade~II~\cite{Blake:2017une}.

\begin{center}
\begin{figure}
    \centering
    \includegraphics[width=0.2\textwidth]{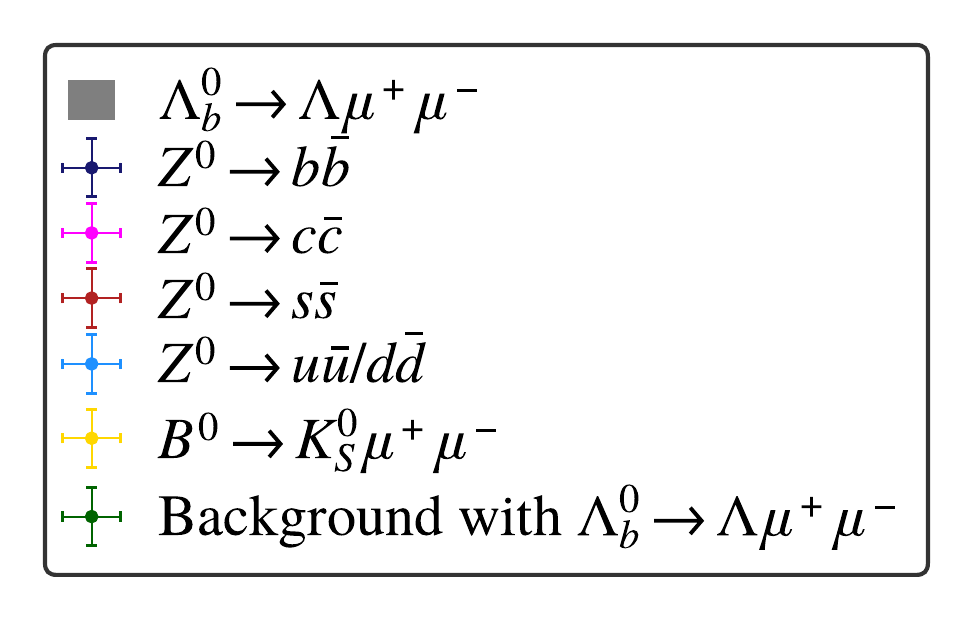}
    \includegraphics[width=0.3\textwidth]{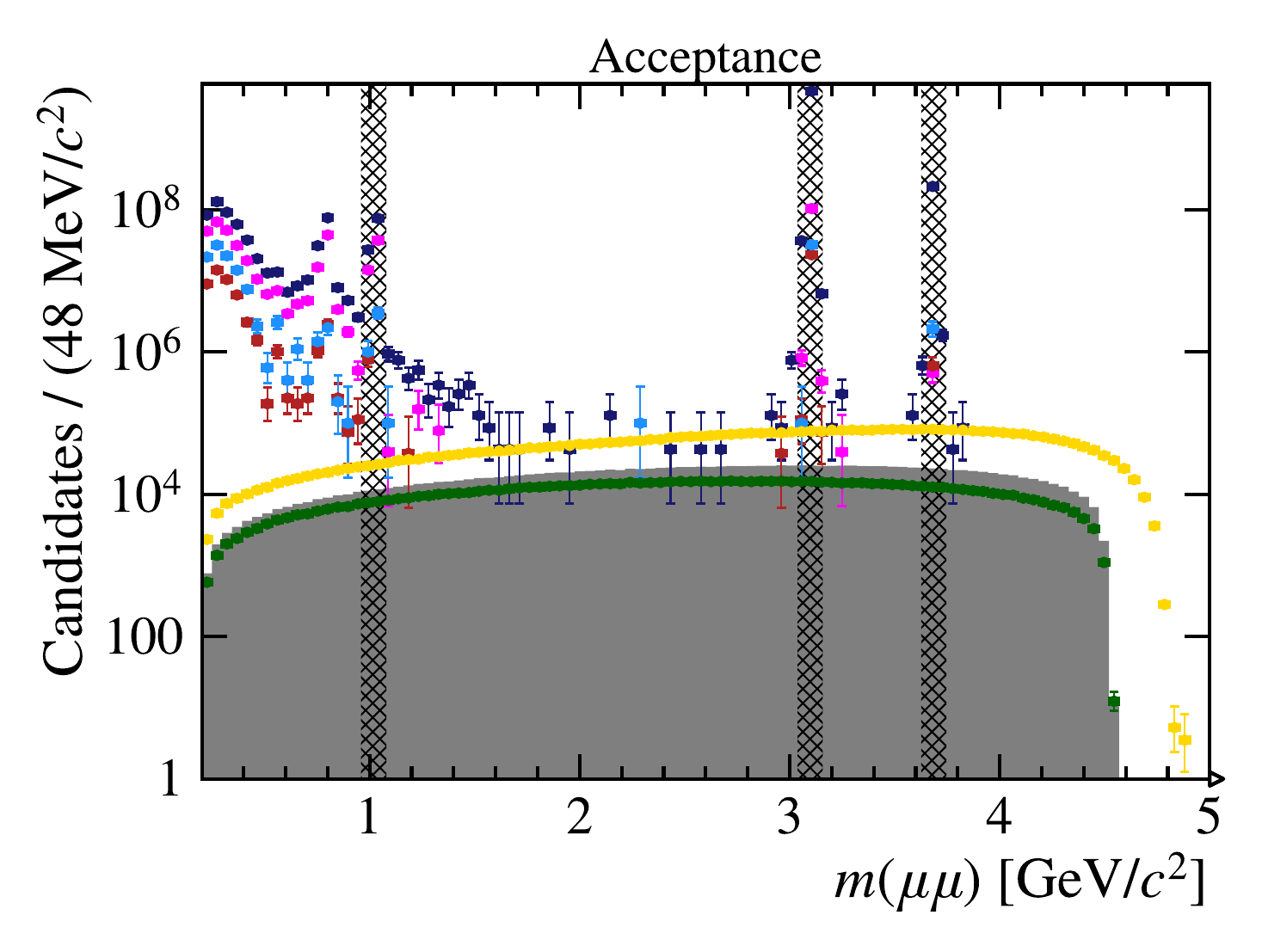}%
    \includegraphics[width=0.3\textwidth]{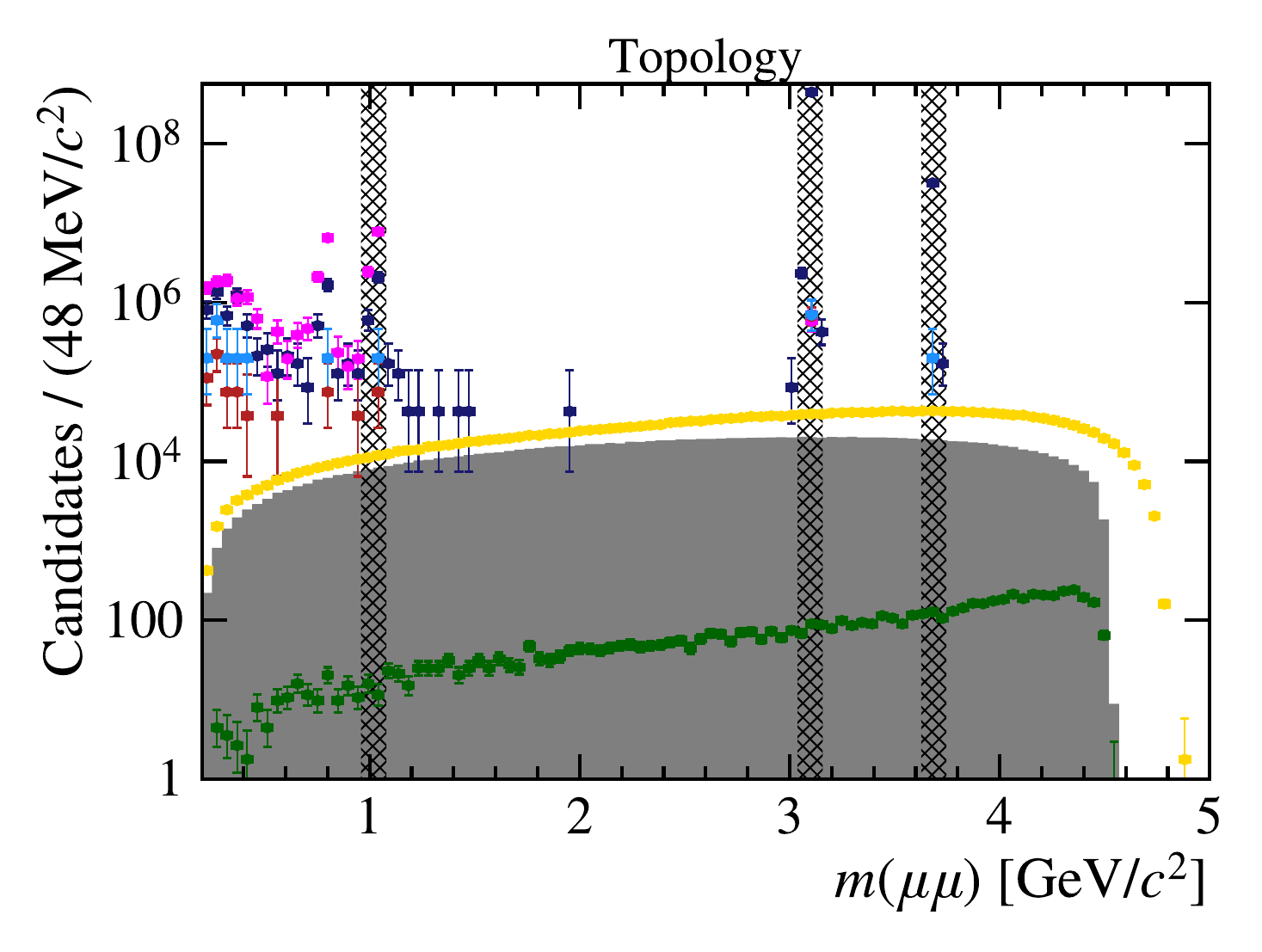}%
    \includegraphics[width=0.3\textwidth]{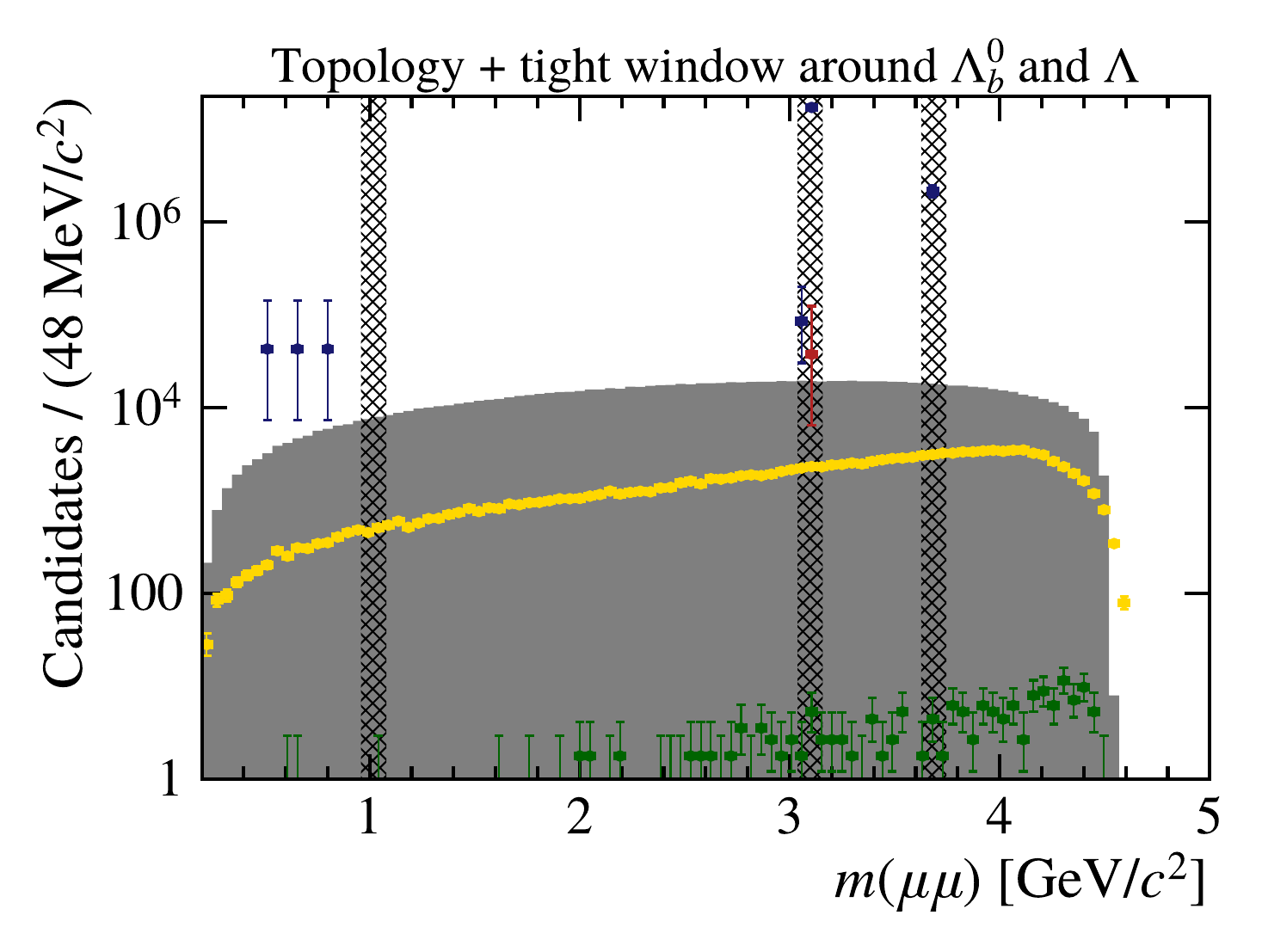}
    \includegraphics[width=0.3\textwidth]{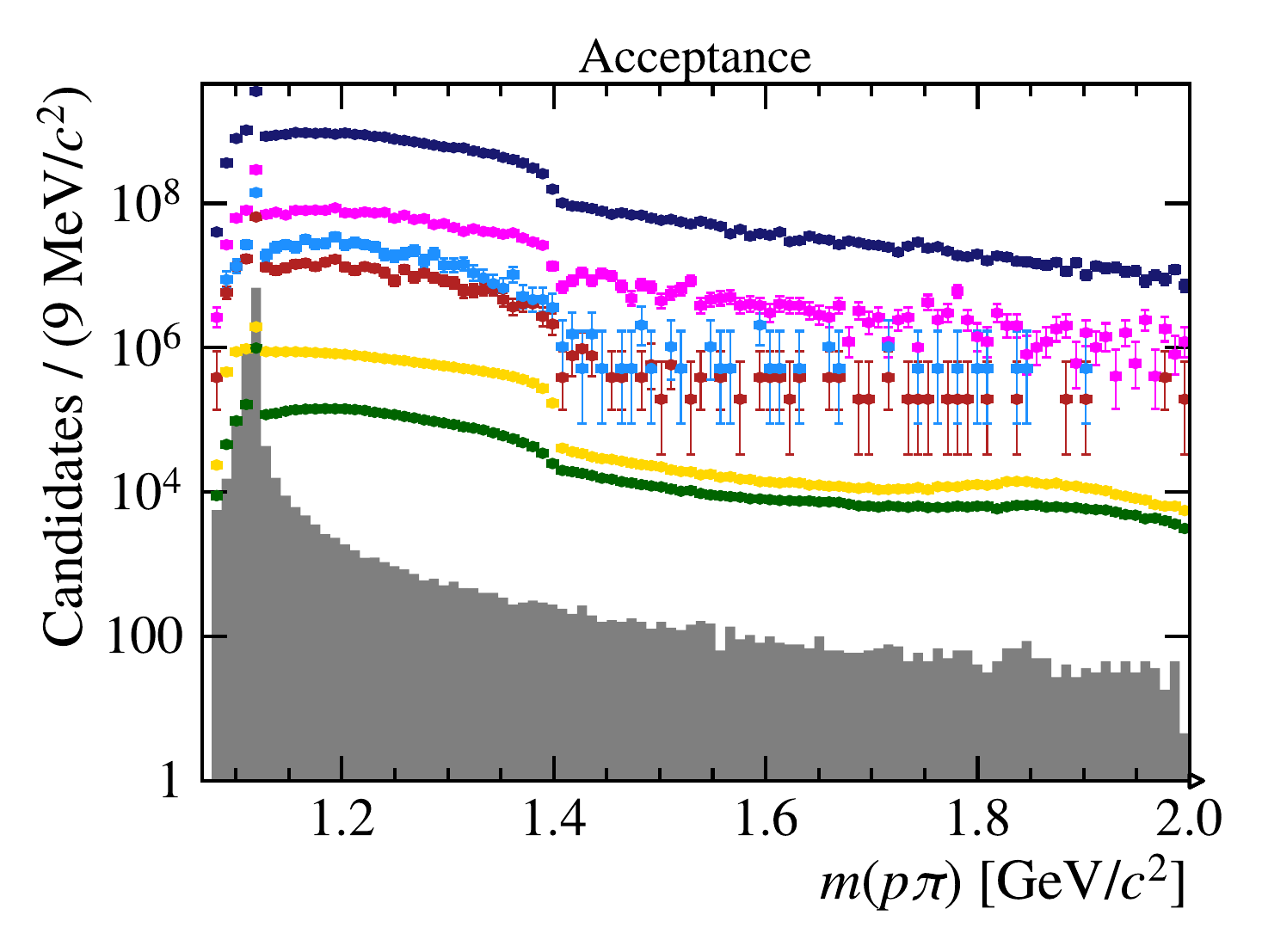}%
    \includegraphics[width=0.3\textwidth]{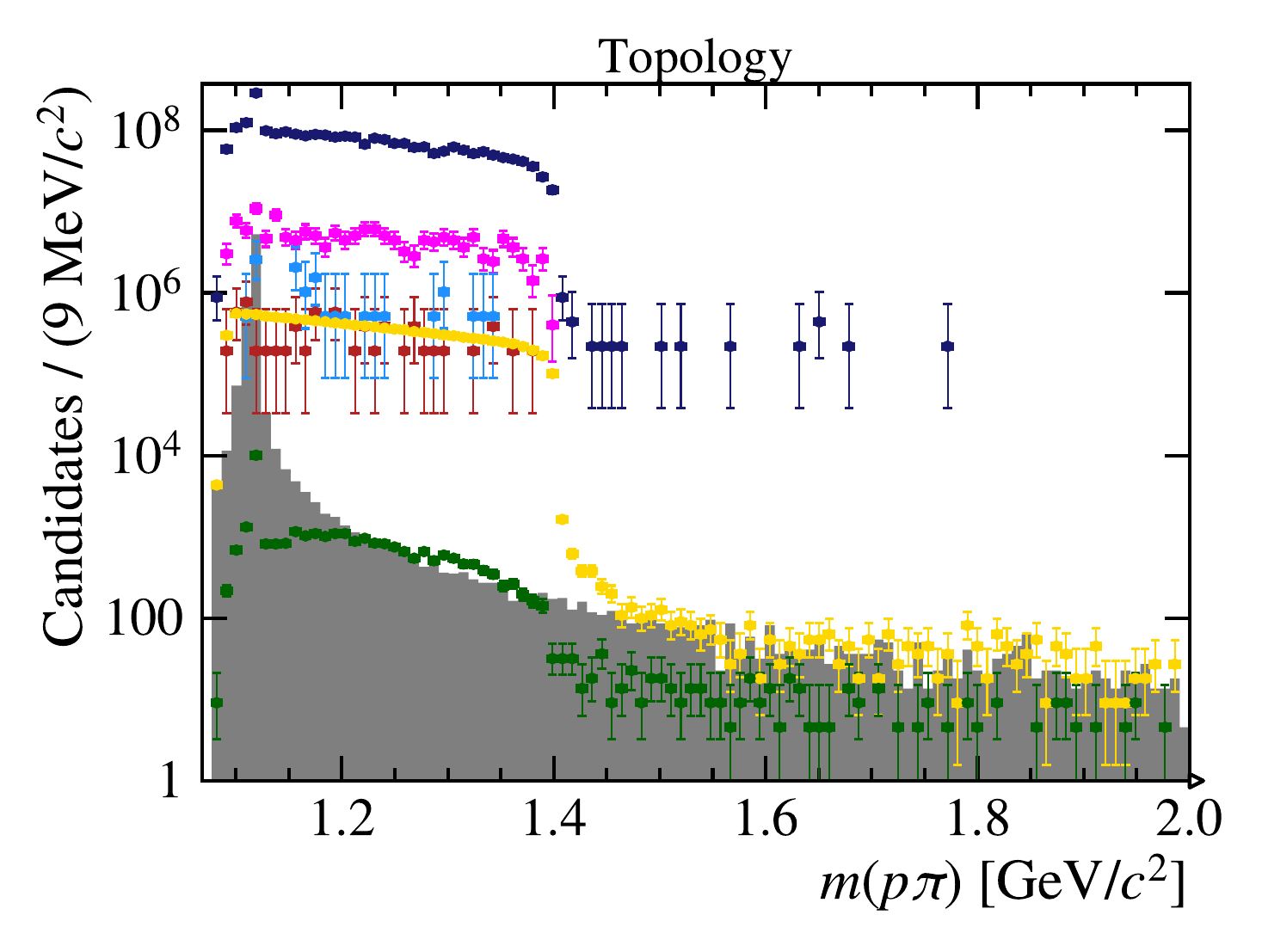}%
    \includegraphics[width=0.3\textwidth]{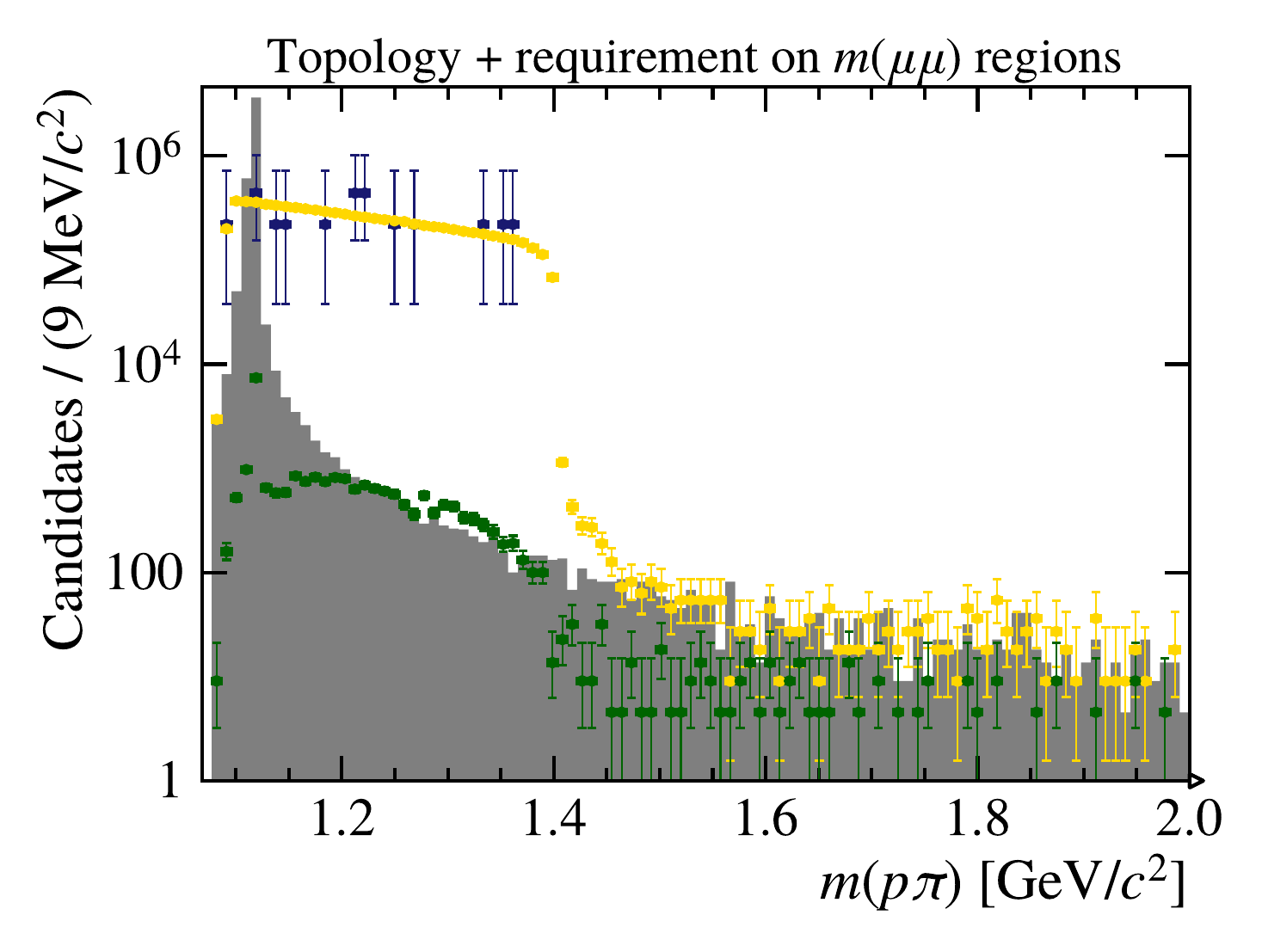}
    \includegraphics[width=0.3\textwidth]{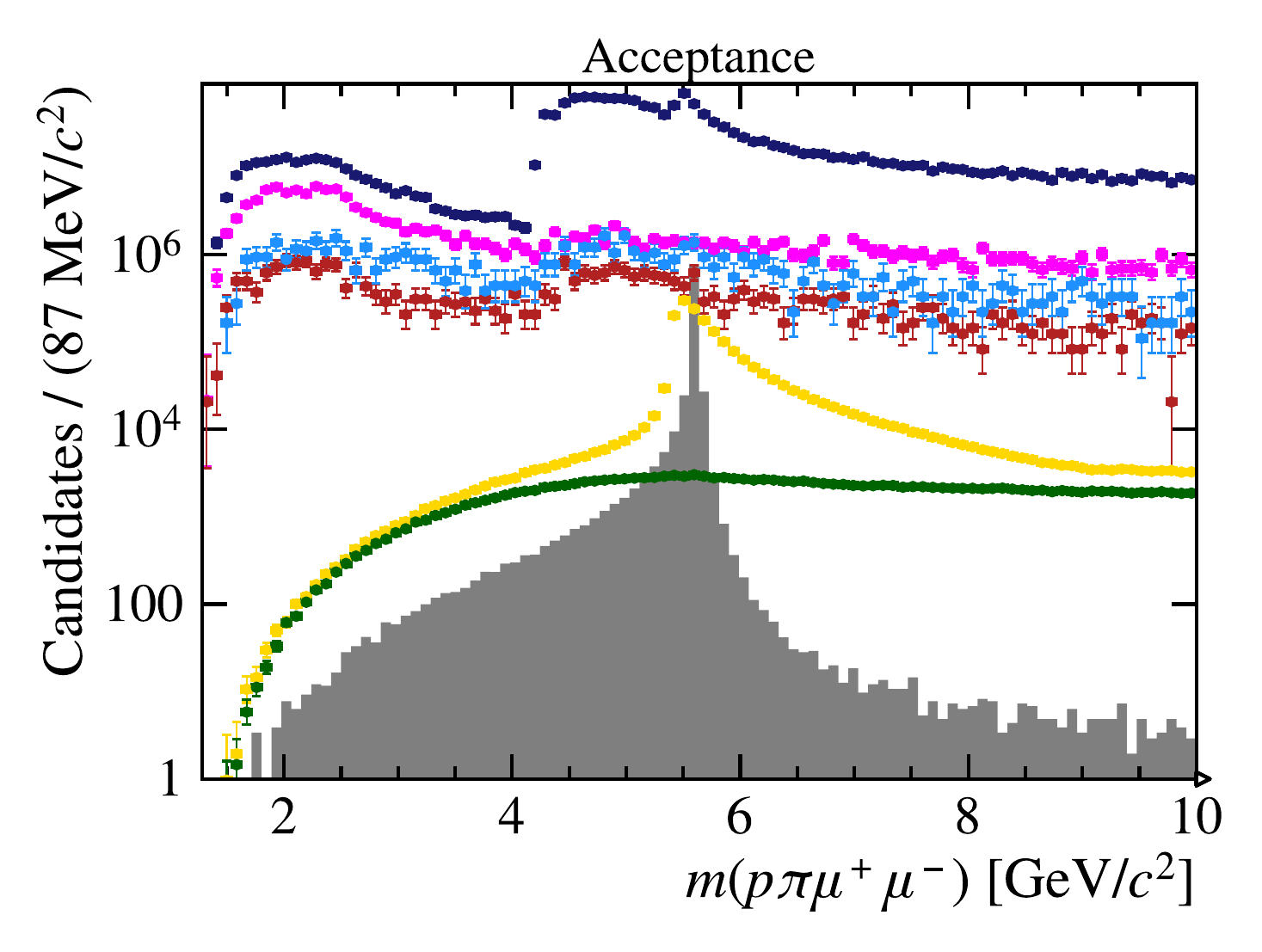}%
    \includegraphics[width=0.3\textwidth]{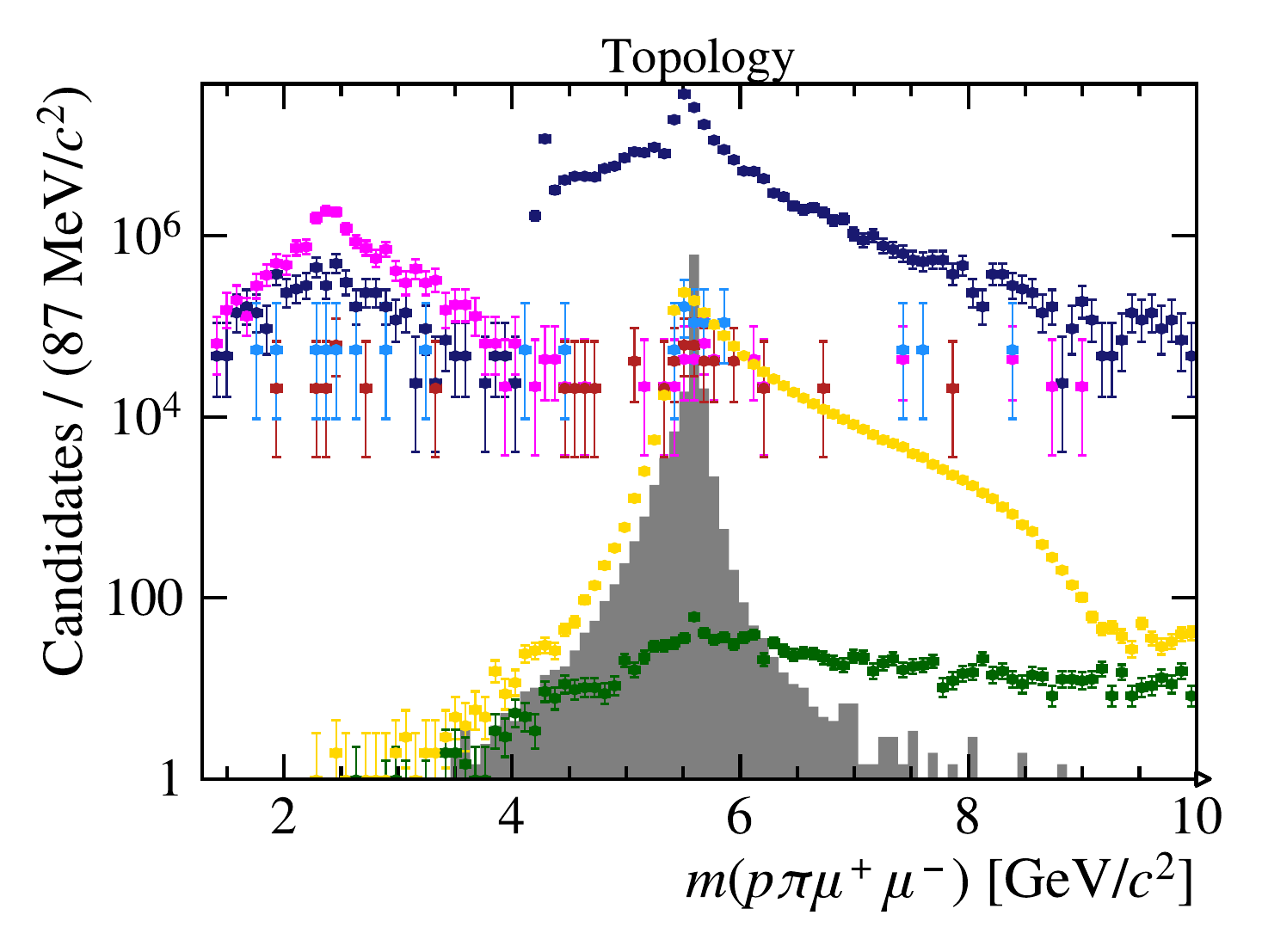}%
    \includegraphics[width=0.3\textwidth]{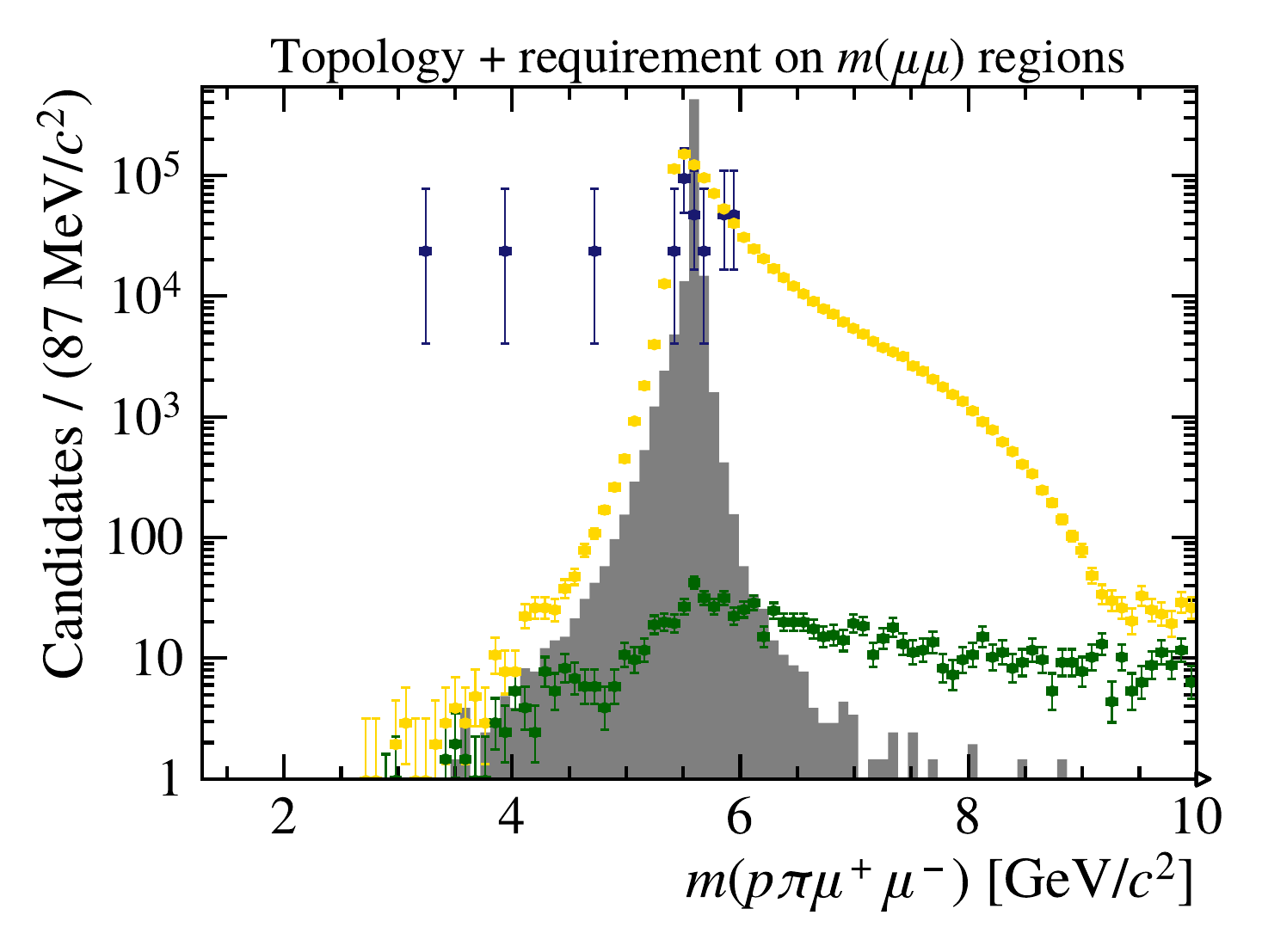}
    \caption{Location and magnitude of the backgrounds (left) after reconstruction without further selections, (center) including the topological, and (right) including all selections (dimuon mass) for the dimuon mass (hadron masses).}
    \label{fig:backgrounds}
\end{figure}
\end{center}

Figure~\ref{fig:backgrounds} displays the distribution of the signal and background samples in the (top) dimuon, (middle) dihadron, and (bottom) four-body invariant masses after different selection steps.
The top row reveals a typical dimuon resonance spectrum in the diquark background samples featuring prominent peaks at the $\phi$, \jpsi, and \psitwos mass.
The region below $1\gevcc$ consists of a more complex combination of resonances.
A study of the simulation samples indicates the presence of dimuon pairs originating from lower mass resonances such as $\eta$ or $\omega(782)$ as well as photons radiated by the \bquark quarks before hadronization.
The middle row shows that some of the dihadron pairs in the background samples are true \Lz baryons peaking at the \Lz mass which are removed by the topological criteria.
The majority however stems from misidentification of one of the hadrons resulting in an almost flat shape similar to the \BdTopipimm background.

\onecolumngrid
\begin{center}
\begin{table}
    \centering
    \caption{Expected magnitudes of potential background sources.
    The second column shows the projected number of produced events assuming $6\cdot10^{12}$ $Z^0$ bosons.
    The columns labeled \textit{Acceptance}, \textit{Topology}, and \textit{Total} show the efficiency for the acceptance, the topological selections, and the requirements on the signal region applied iteratively.
    The efficiency is calculated as the number of reconstructed candidates divided by the number of generated events, which can lead to values larger than unity as the reconstruction may give up to one \Lb and one \Lbbar candidate per event.
    The rightmost column contains the projected number of candidates after selection.
    If no candidates remain after a given selection, the efficiency or number is given as an upper limit calculated as the 68\% confidence interval.
    }
    \label{tab:contaminations}
    \input{contaminations}
\end{table}
\end{center}
\twocolumngrid

The bottom row reveals some interesting structures appearing in the four-body invariant mass even after enforcing the correct topology.
The topological selections remove the background from light quarks almost entirely.
The \ccbar and \bbbar backgrounds appear in similar shapes just above 2\gevcc.
This accumulation is the result of combining a low-mass dimuon resonance produced in a \bquark- or \cquark decay combined with a dihadron pair produced in the hadronization.
These prompt hadrons have very low momentum leading to the low four-body invariant mass.
The displacement of the dimuon vertex from the primary vertex due to the long \bquark- and \cquark-quark lifetimes allows these combinations to pass the topological selections.
Additionally, the \bbbar background distribution peaks around the \Lb mass.
This peak stems from \Lb decays to a \Lz baryon and a \qqbar meson as well as similar \Bd decays with misidentified hadrons.
The very narrow peak in the \bbbar sample at 4.25\gevcc originates from $\Lb\to \jpsi(\to\mumu)\Lz(\to p\pim)\phi$ decays.
Despite not reconstructing the $\phi$ meson, this decay still leads to a peaking structure due to the narrow \jpsi and \Lz resonances as well as the very limited phase space remaining after producing the $\phi$, \jpsi, and \Lz mesons.
The branching fraction of $\Lb\to\jpsi\Lz\phi$ is 4\% of the branching fraction of $\Lb\to\jpsi\Lz$~\cite{PDG2024}, which is roughly consistent with the magnitude of the peak.

At first glance, the upper limits on the selected background candidates in the $Z^0\to\qqbar$ samples may appear unsatisfying.
They are however a direct consequence of the number of generated events lying between 300 and 500 million for each species, three orders of magnitude below the expected numbers that will be produced over the full $Z$-pole run at the FCC-ee, see Table~\ref{tab:contaminations}.
A reduction of the upper limit on the total contamination due to $Z^0\to\qqbar$ events to below 1\% of the signal yield at 68\% confidence level would require a 50-fold increase in simulation samples which is not feasible.
Given the discussions above however, the light-quark $Z^0\to\qqbar$ backgrounds exclusively populate the low dimuon invariant mass region not considered in the analysis while the heavier quark backgrounds also sit at dimuon \qqbar resonance regions which are vetoed.
As a consequence, there is high confidence that the backgrounds can be neglected given the definition of the signal region and analysis bins.

\section{Relative efficiency}
The decay of a spin-1 boson to two fermions---such as $Z^0\to\bbbar$---is not isotropic leading to decay products with preference for large pseudorapidity.
Moreover, the momentum distribution of \Lb baryons produced in $Z^0$ decays peaks between 35 and 40\gevc.
This strong boost causes the decay products to fly in a narrow cone around the \Lb momentum in the laboratory frame.
The beam line however limits the acceptance of the detector to tracks whose angle with respect to the beam axis is $\alpha\geq\arccos(0.99)$.
This design detail in combination with the production kinematics limits the integrated acceptance for \LbToppimm decays in this study to around 80\%. 
Another challenge is the reconstruction of low-momentum particles.
The angles between the particles dictate the magnitude of the momentum after the boost to the lab frame leading to nonuniform relative efficiencies.
The most prominent effects and their causes are discussed later in this Section.

The efficiency in the six-dimensional phase space \mbox{$\vec{\phsp}=(\qprime,\cos\tp,\cos\tm,\cos\th,\phim,\phih)$} is parametrized by
\begin{widetext}
\begin{equation}
    \varepsilon(\vec{\phsp}) = \sum\limits_{n_q n_\parallel n_\mu n_h l_\mu l_h} e_{n_q n_\parallel n_\mu n_h l_\mu l_h} P_{n_q}(\qprime)P_{n_\parallel}(\cos\tp)P_{n_\mu}(\cos\tm)P_{n_h}(\cos\th) \cos (l_\mu\phim)\cos (l_h\phih)~,
\end{equation}
\end{widetext}
where $P_i(x)$ are Legendre polynomials of order $i$, the $e_{n_q n_\parallel n_\mu n_h l_\mu l_h}$ are a set of coefficients, and \qprime is a mapping of $\sqrt{\qsq}$ to the range $[-1,+1]$. 
This model makes no assumptions about the factorization of the different observables. 
Legendre polynomials, and $\cos m\phi$ dependencies, are used since these form an orthogonal basis.
Note that for the $\cos\tm$ and $\phim$ dependence only even orders are allowed as there is no reason for a detection asymmetry on the muonic side of the decay.
The coefficients $e_{n_q n_\parallel n_\mu n_h l_\mu l_h}$ are determined using the method of moments on phase space simulation samples. 
The efficiency is parametrized using polynomials of up to and including order 2 for the transformed mass, \qprime, order 7 for $\cos\tp$, order 0 for $\cos\tm$, order 2 for $\cos\tp$, order 4 for \phim, and order 2 for \phih.
These orders are chosen based on a number of criteria outlined in the following.

\begin{center}
\begin{figure}
    \centering
    \includegraphics[width=\linewidth]{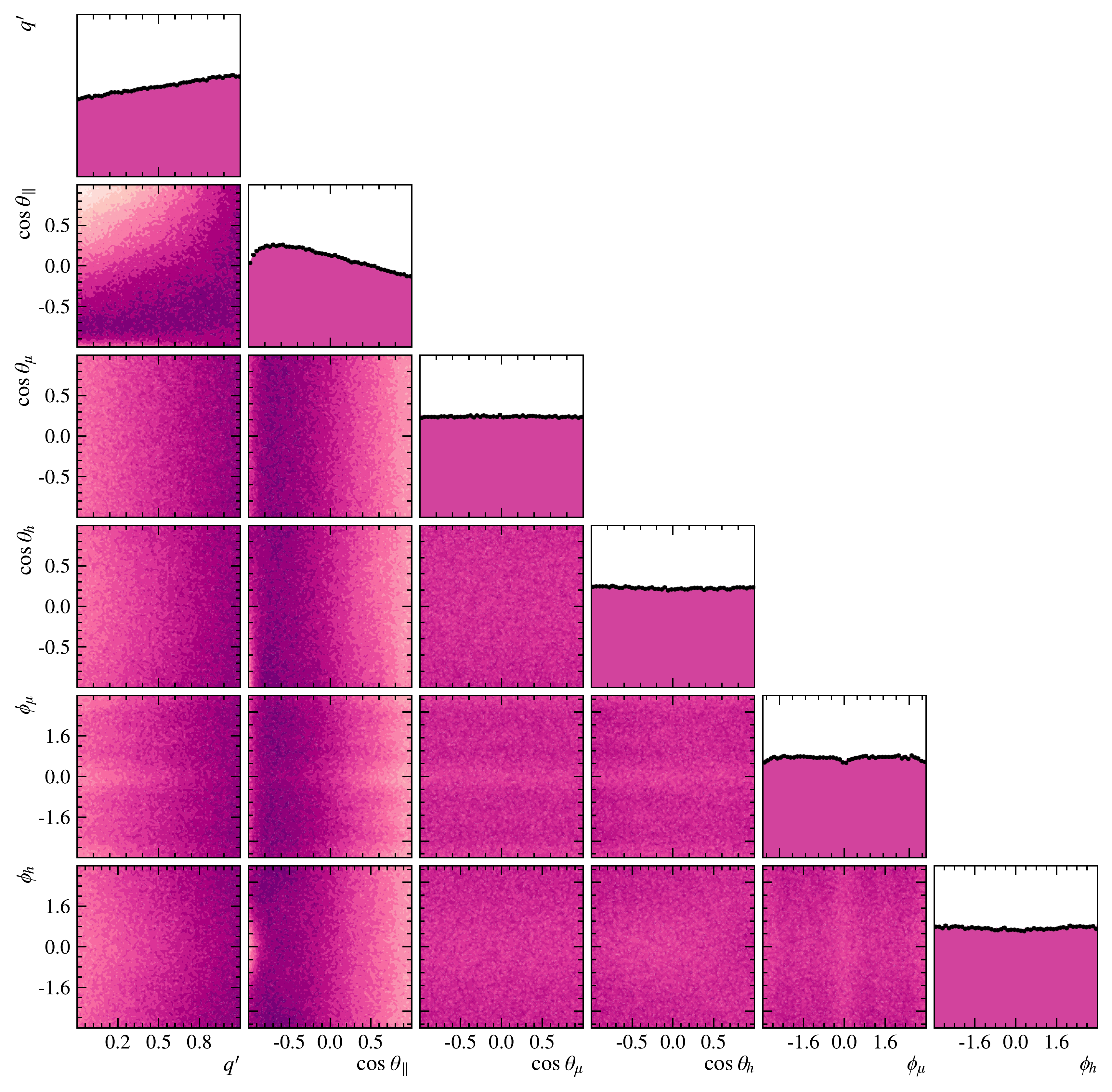}
    \caption{One- and two-dimensional projections of the efficiency model.
    For the one-dimensional projections, the simulation samples (including calibration and after removal of phase space structures) are displayed as black data points.
    The colors in all off-diagonal plots are have the same scale.}
    \label{eff:fig}
\end{figure}
\end{center}

The diagonal plots in Fig.~\ref{eff:fig} show the one-dimensional efficiency projections.
These shapes provide a guide for the minimum orders required in the model such as a linear term for \qprime and high orders in $\cos\tp$ and $\phim$.
The efficiency model with the chosen orders (pink shape) agrees well with the true efficiency represented by the simulation samples (black histogram).
The off-diagonal two-dimensional projections show the efficiency model where the colors are scaled consistently across all combinations of variables to illustrate the large changes across \qprime and $\cos\tp$ compared to all other variables.

\onecolumngrid
\twocolumngrid

\begin{figure}
    \centering
    \includegraphics[width=.9\linewidth]{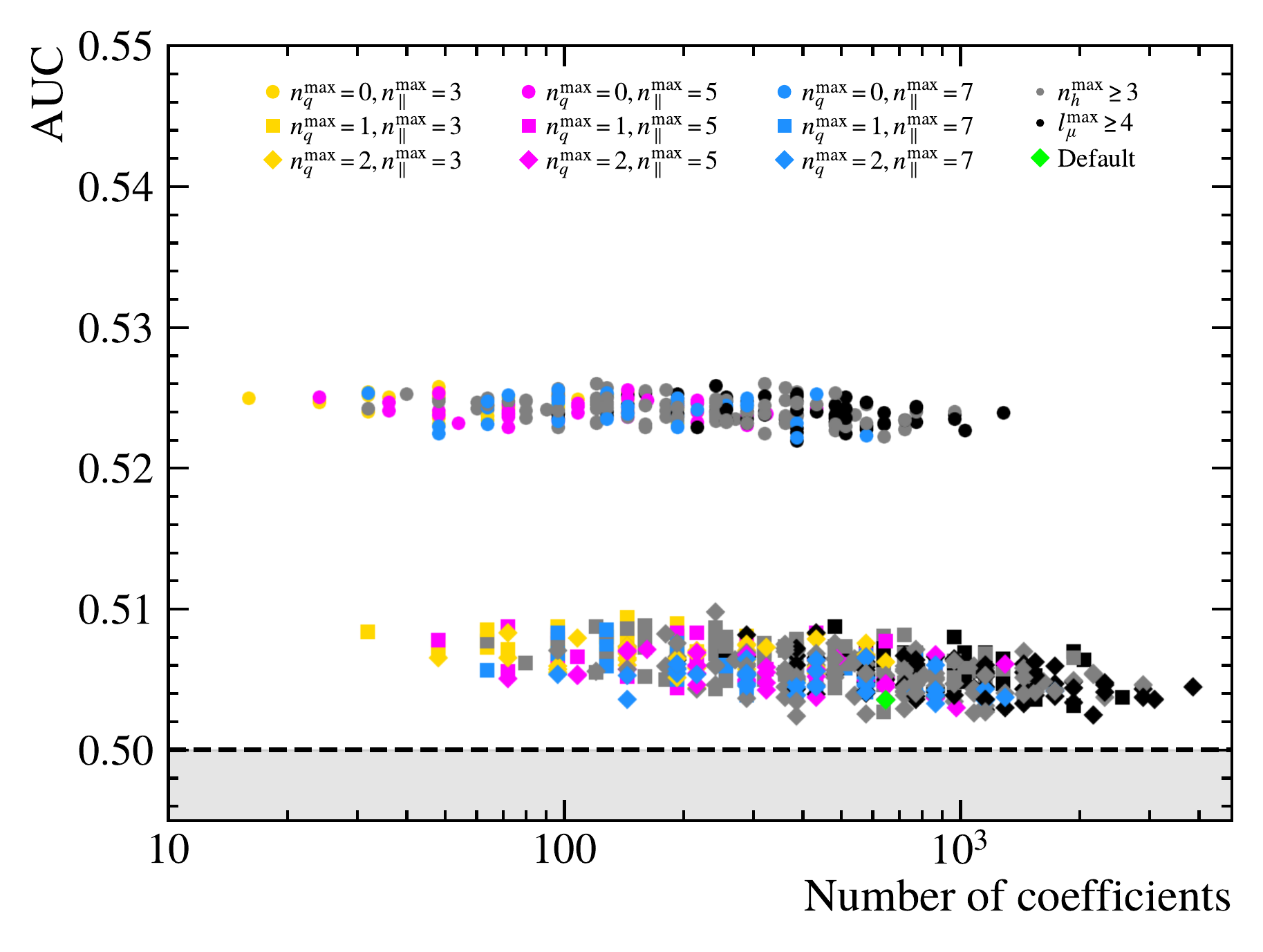}
    \caption[Validation of the efficiency mode.]{%
    AUC for classifying a toy of an efficiency model from the simulation samples.
    The nominal model is shown in green.
    The dashed line at 0.5 represents inseparable samples.
    Values below 0.5 are impossible and indicate a problem with the classifier or the input data.
    The horizontal axis indicates the complexity of the model given by the number of coefficients.}
    \label{eff:fig:validation}
\end{figure}

\clearpage
In order to study the agreement quantitatively and including correlations across all six dimensions, a large toy sample representing the efficiency model is generated.
A Boosted Decision Tree (BDT) with the default \mbox{\textsc{XGBoost}}~\cite{Chen_2016} configuration is trained employing a twofold strategy to distinguish the simulation and toy samples.
Its ability to disentangle them is quantified using the area under the receiver-operator curve (AUC) obtained when classifying the training samples.
Figure~\ref{eff:fig:validation} shows the AUC for efficiency models covering reasonable combinations of orders.
The horizontal axis represents the complexity of the model quantified by the number of coefficients.
The different marker shapes represent the different orders in \qprime.
A uniform efficiency in \qprime, labeled by $n_q\leq0$, disagrees with the one-dimensional projection shown in Fig.~\ref{eff:fig}.
This option is included in the plot however to illustrate that obvious deviations from the correct model appear as significant differences in AUC whereas higher-order improvements, such as going from $n_q\leq1$ to $n_q\leq2$ have much smaller impact.
The different colors represent different orders in $\cos\tp$ showing no clear separation between different orders.
Similarly models with $l_\mu^\text{max}\geq4$ are shown in black and models with $n_h^\text{max}\geq3$ are shown in gray again with no obvious improvement over other colors.
While the general trend is that AUC improves slowly with the complexity of the model, there is no drastic improvement related to any variable.

Only models with variable importances in the classification that lie within $\tfrac{1}{6}\pm0.010$,  and which have no regions with negative efficiency,  are considered as potential efficiency parametrizations.
Requiring very similar variable importances chooses models that have no strong mismodeling in one variable.
Moving to higher orders than tried in this test will lead to an even lower AUC.
It however also increases the risk of localized negative efficiency values which the BDT cannot penalize because a toy sample can never represent a negative density.
Amongst the preselected models, the one with the lowest AUC score---of 0.5035---is chosen as the default.
In conclusion, the default efficiency model is almost indistinguishable from the simulation samples and, given the limitations of the validation method, no efficiency model with significantly better performance can be identified.

Returning to the efficiency shapes in Fig.~\ref{eff:fig}, the locally lower efficiencies are due to lower momenta in one of the particles.
The momentum of the proton and the pion is equal in the \Lz rest frame but the proton has significantly higher energy.
This higher energy leads to on average higher proton momentum after a boost to another frame.
The pion momentum is particularly low when the combined boost of the \Lb and \Lz is antiparallel to the pion momentum in the \Lz rest frame.
This happens for a low-momentum hadronic system when the \Lz in the \Lb rest frame flies antiparallel to the \Lb lab momentum ($\cos\tp\approx-1$), the pion in the \Lz rest frame flies antiparallel to the \Lb lab momentum ($\phih\approx0$), and the pion is emitted parallel to the \Lz ($\cos\th<0$).
The same happens for a low-momentum dimuon system when the \Lz in the \Lb rest frame flies parallel to the \Lb lab momentum ($\cos\tp\approx1$), and either the muon of the same charge as the proton flies antiparallel to the \Lz ($\cos\th<0$) and antiparallel to the \Lb lab momentum ($\phim\approx\pm\pi$), or  the muon of the opposite charge as the proton flies antiparallel to the \Lz ($\cos\tm>0$) and antiparallel to the \Lb lab momentum ($\phim\approx0$).

\section{Angular fit}
The angular observables $K_i$ of \LbToppimm are determined using a maximum likelihood fit of the five-dimensional angular decay rate given in Eq.~\eqref{eq:decayrate}.
The background levels are negligible and for the nominal fit result, particle identification is assumed to be perfect.
As a consequence, there are no additional shapes in the fit.
The fit is repeated on 1000 toy samples of realistic size.
An examination of the pull distribution confirms unbiased results and accurate coverage of the Hessian uncertainties.
Similar studies for different values of the polarization show compatible uncertainties and reliable fit stability confirming the robustness of the setup.
As an additional consistency check, it has been verified that the values for the angular coefficients obtained from the fit are in agreement with the values calculated using the method of moments, which is by construction model independent.
Moreover, we confirm that the extracted observables fulfil the theoretical relationships $K_{15} = -\alpha_\Lambda P_\parallel K_2$ (which is only true in the limit of massless leptons), $\alpha_\Lambda K_{13}=-P_\parallel K_6$, and $K_{16}=-\alpha_\Lambda P_\parallel K_3$, with the $\Lambda$ decay asymmetry parameter $\alpha_\Lambda$~\cite{Blake:2017une}.

The systematic uncertainty is calculated as the difference between the true generated value and the measured value on the full simulation sample after the full selection.
This value includes all potential sources of systematic uncertainty at once and is assessed for different levels of contamination due to misidentified \BdTopipimm decays.
For a $p-\pi$ separation of $\kappa\sigma$, a fraction
\begin{align}
    f = \frac{1}{2}\left(1-\text{cdf}_{\chi^2}(\kappa^2,1)\right)
\end{align}
of the signal is replaced by \BdTopipimm samples, where $\text{cdf}_{\chi^2}(x,d)$ is the cumulative distribution function of a $\chi^2$ distribution with $d$ degrees of freedom.
As discussed previously, the efficiency is included by weighting the samples in the fit.
Given the reconstruction and selection focuses on the decay geometry, the efficiency for \BdTopipimm is assumed to be very similar to the signal efficiency and no specific efficiency model is developed.

\begin{figure}
    \centering
    \includegraphics[width=.9\linewidth]{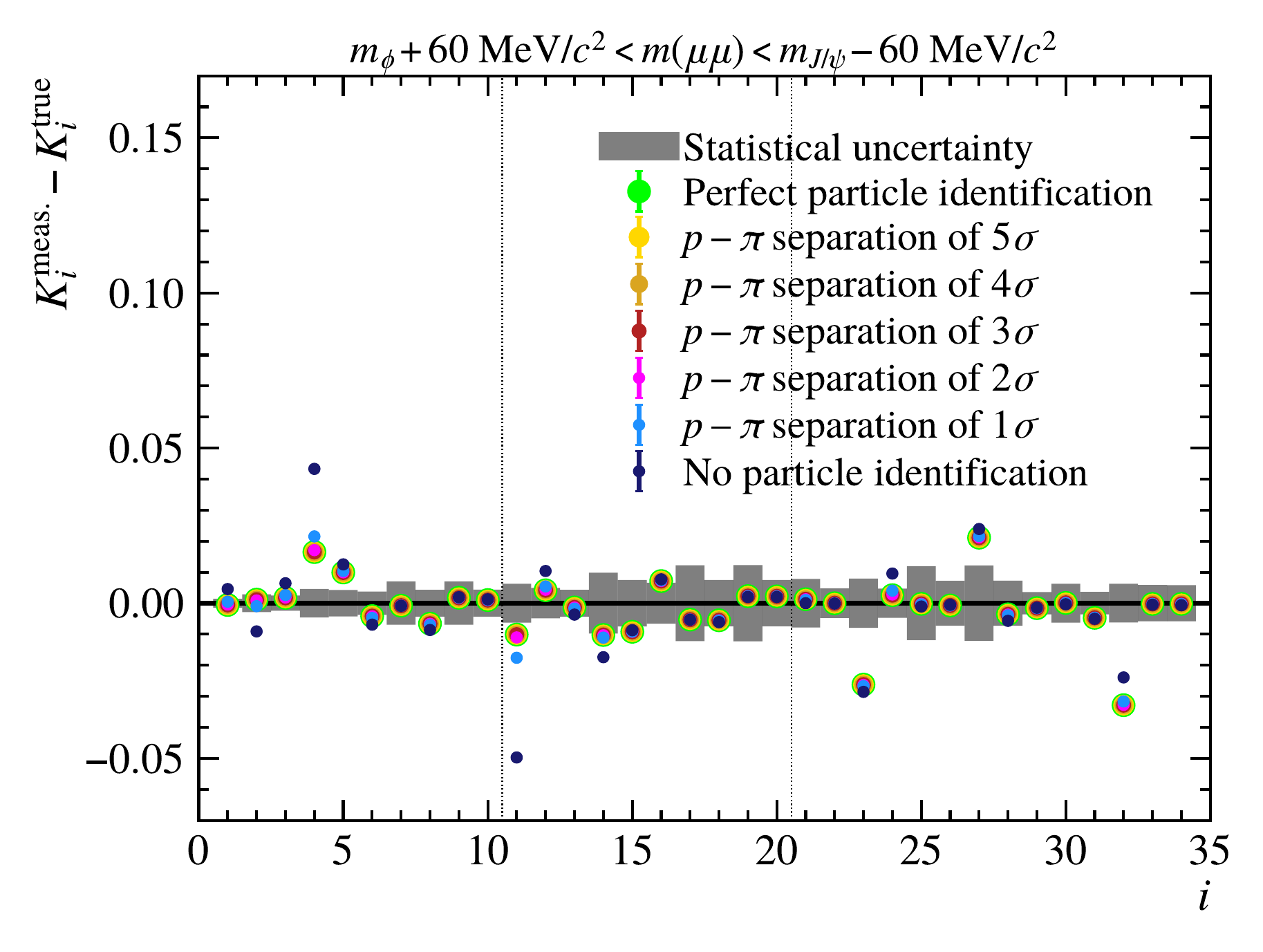}
    \includegraphics[width=.9\linewidth]{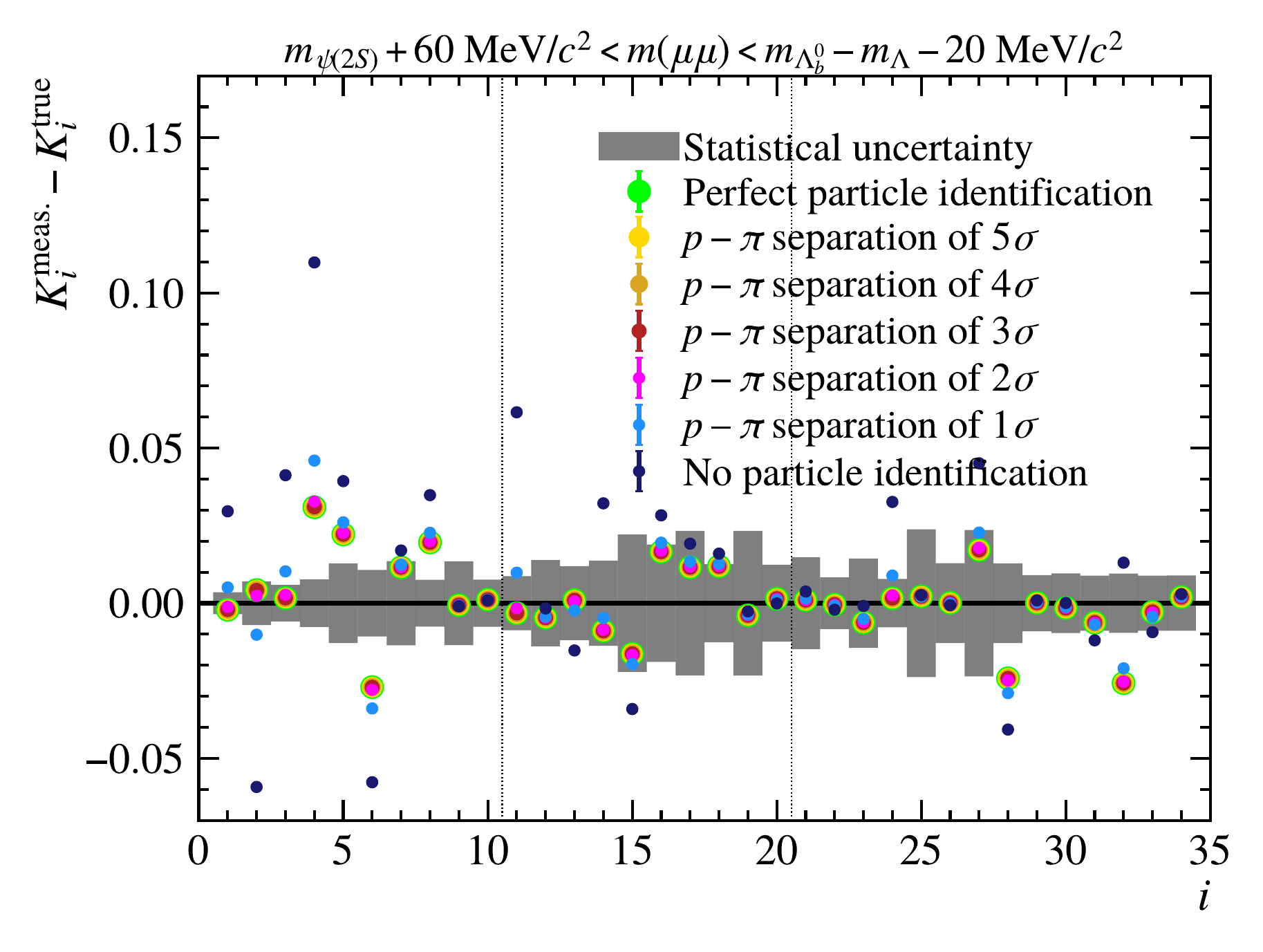}
    \caption{Uncertainties in the measurement for different levels of $p-\pi$ separation obtained as the deviation of an observable from the truth.
    }
    \label{fig:systematic}
\end{figure}

Figure~\ref{fig:systematic} shows the statistical uncertainty as gray bands and the total systematic uncertainty represented as the deviation from the true value for different levels of $p-\pi$ separation.
Table~\ref{tab:observables} in Appendix~\ref{app:observables} summarizes the numerical results.
The uncertainties are slightly larger than the projected uncertainties for a moment analysis in the region of large dimuon invariant mass performed by the LHCb experiment after Upgrade~II.
This is within expectation given the significantly lower yield in this study.
Some loss of sensitivity is however recovered by performing an angular fit as opposed to the method of moments employed in the LHCb projection.
Most observables are statistically limited while some have larger systematic uncertainty which remain even in the case of perfect particle identification, in particular $K_{4,5,6,14,23,32}$.
Beyond particle misidentification, the largest sources of systematic uncertainty are assumed to be a mismodeling of the efficiency and resolution effects.
The systematic uncertainty due to faulty particle identification practically vanishes beyond a moderate separation of $2\sigma$ for all observables.

\section{Sensitivity to the Wilson coefficients \texorpdfstring{$C_{9^{(\prime)}}$}{C9} and \texorpdfstring{$C_{10^{(\prime)}}$}{C10}}
\begin{figure}
    \centering
    \includegraphics[width=0.8\linewidth]{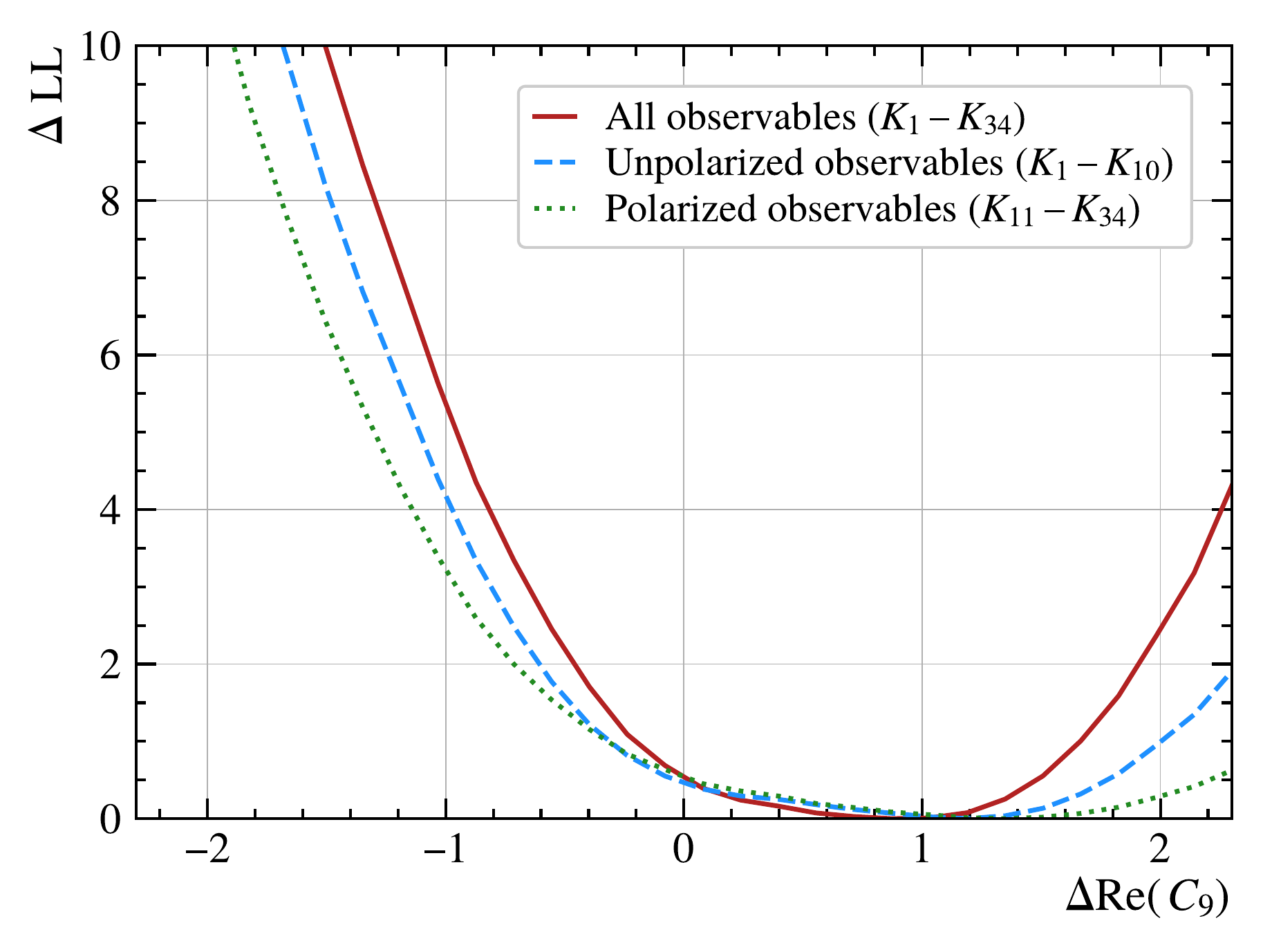}
    \includegraphics[width=0.8\linewidth]{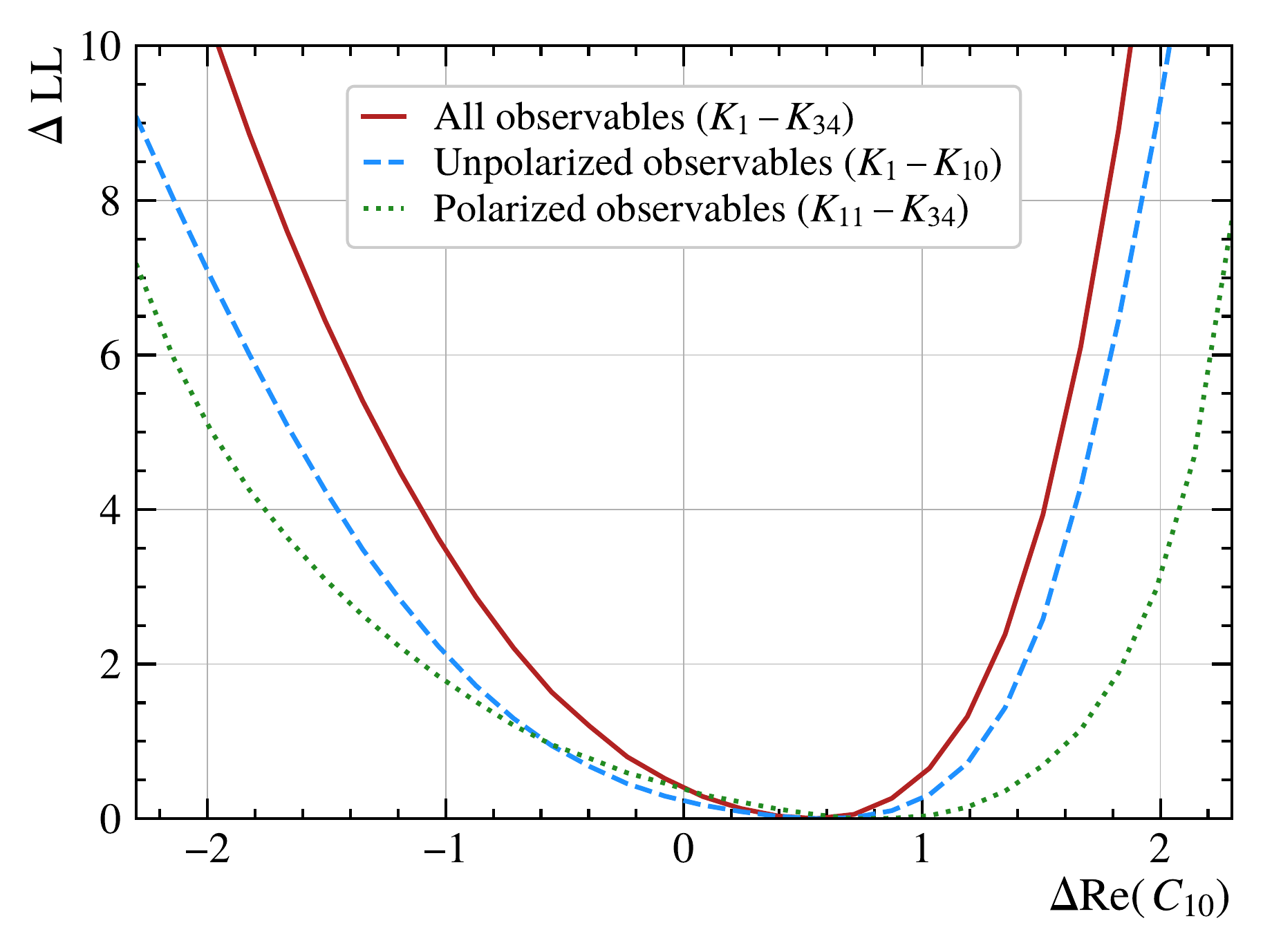}
    \includegraphics[width=0.8\linewidth]{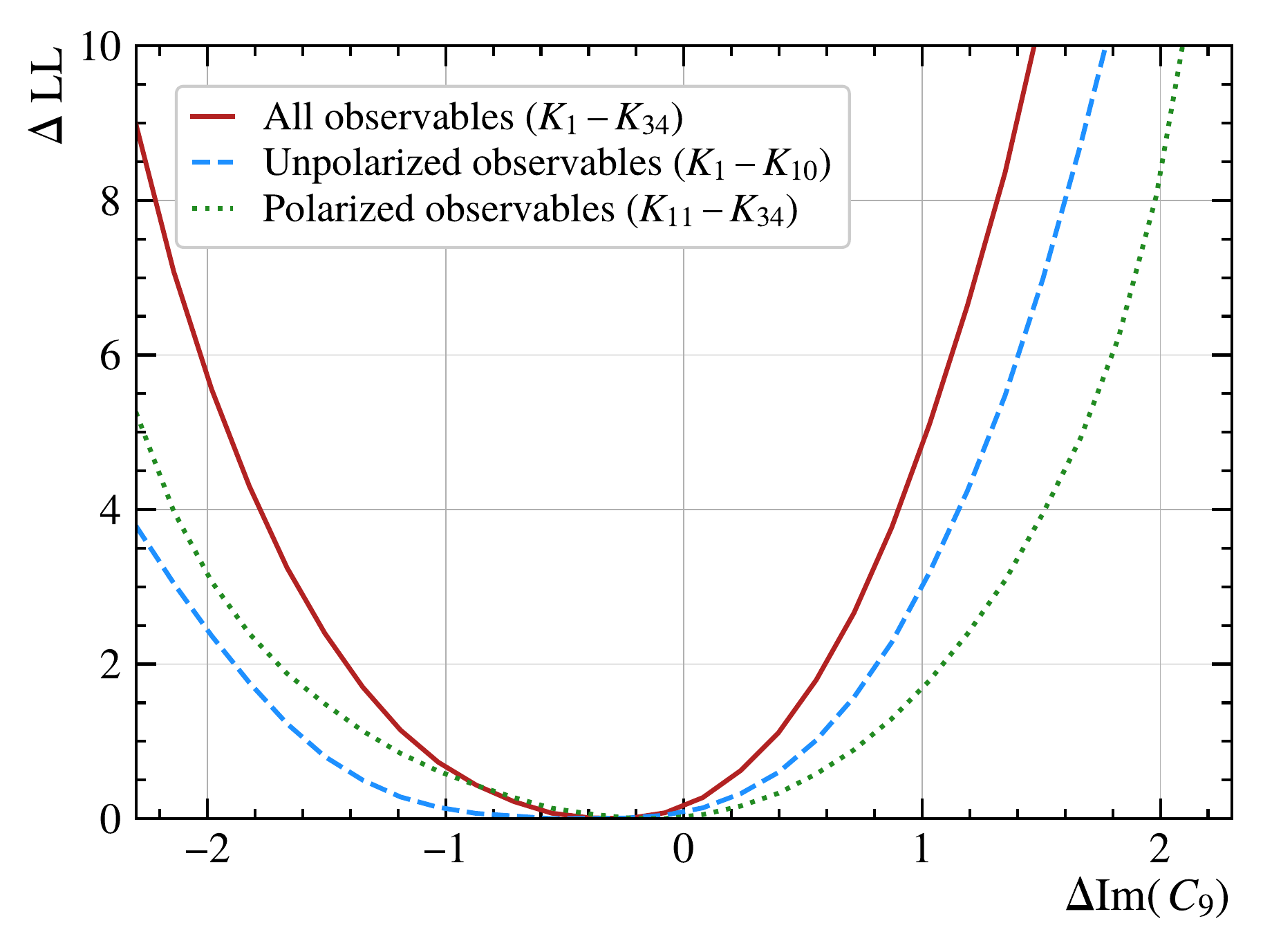}
    \includegraphics[width=0.8\linewidth]{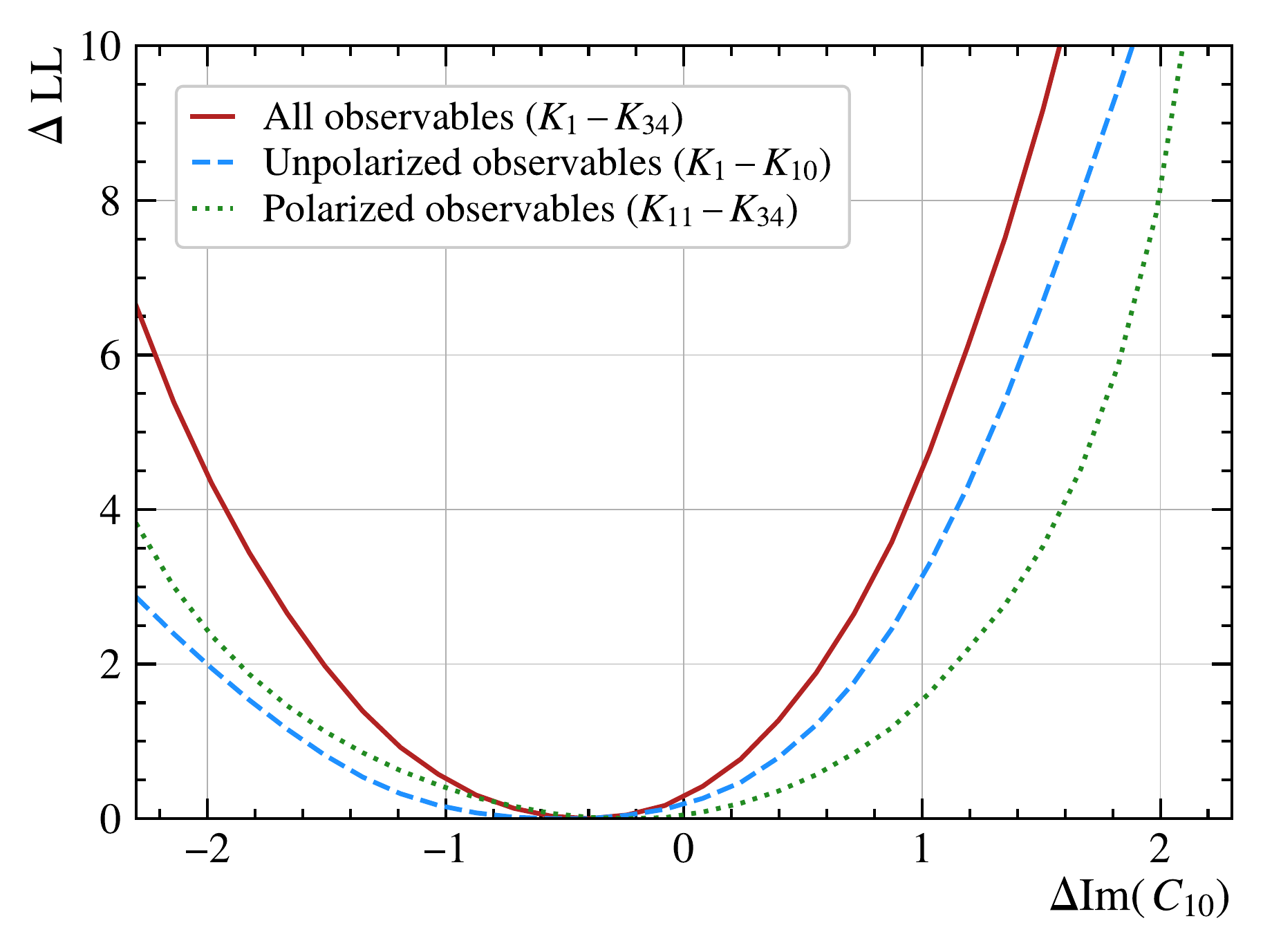}
    \caption{Sensitivity of a fit to the real and imaginary parts of the Wilson coefficients $C_9$ and $C_{10}$ for a measurement using either all observables, only unpolarized, or only polarized observables assuming perfect particle identification.}
    \label{fig:sensitivity:noprime}
\end{figure}

\begin{figure}
    \centering
    \includegraphics[width=0.8\linewidth]{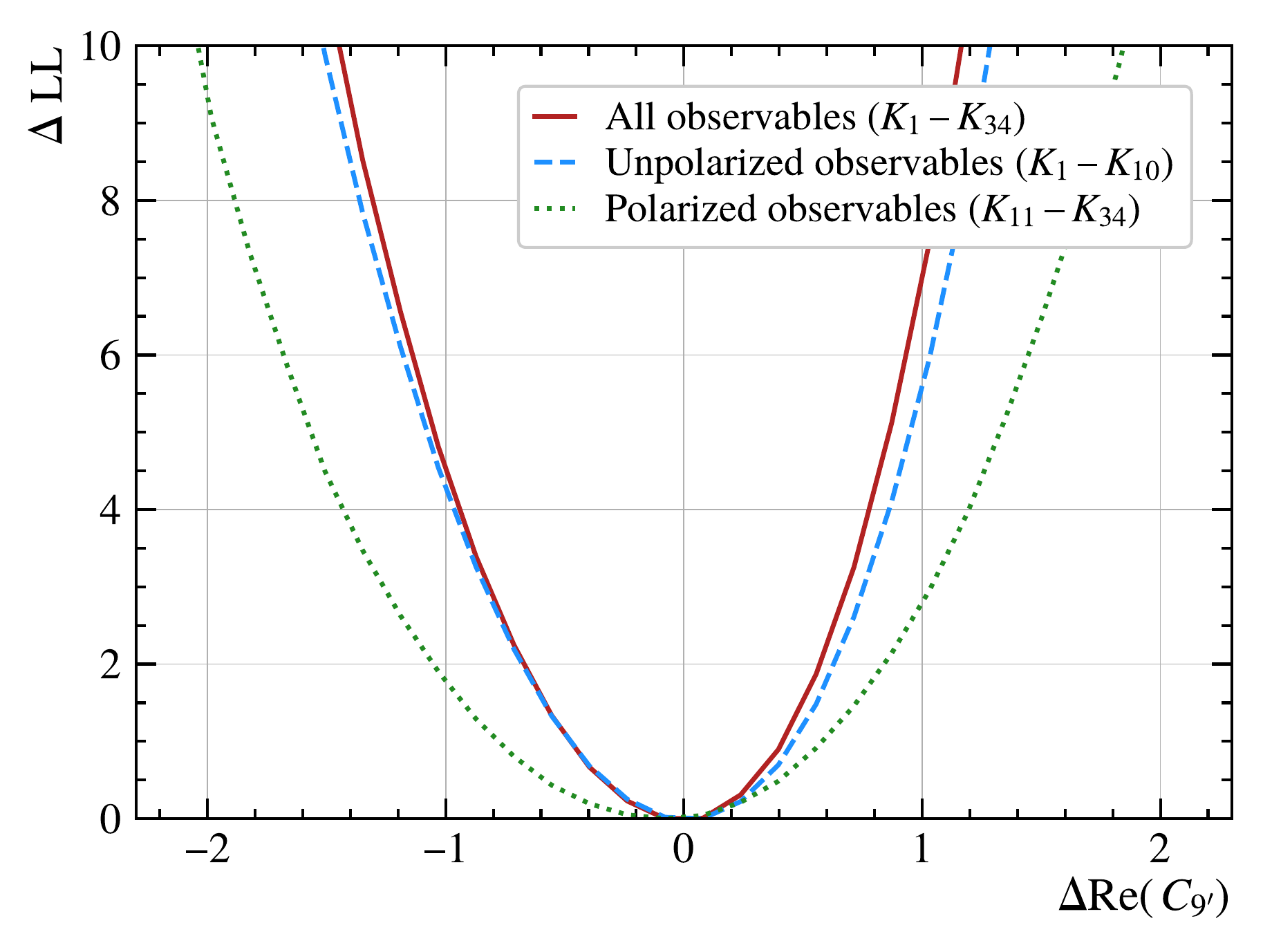}
    \includegraphics[width=0.8\linewidth]{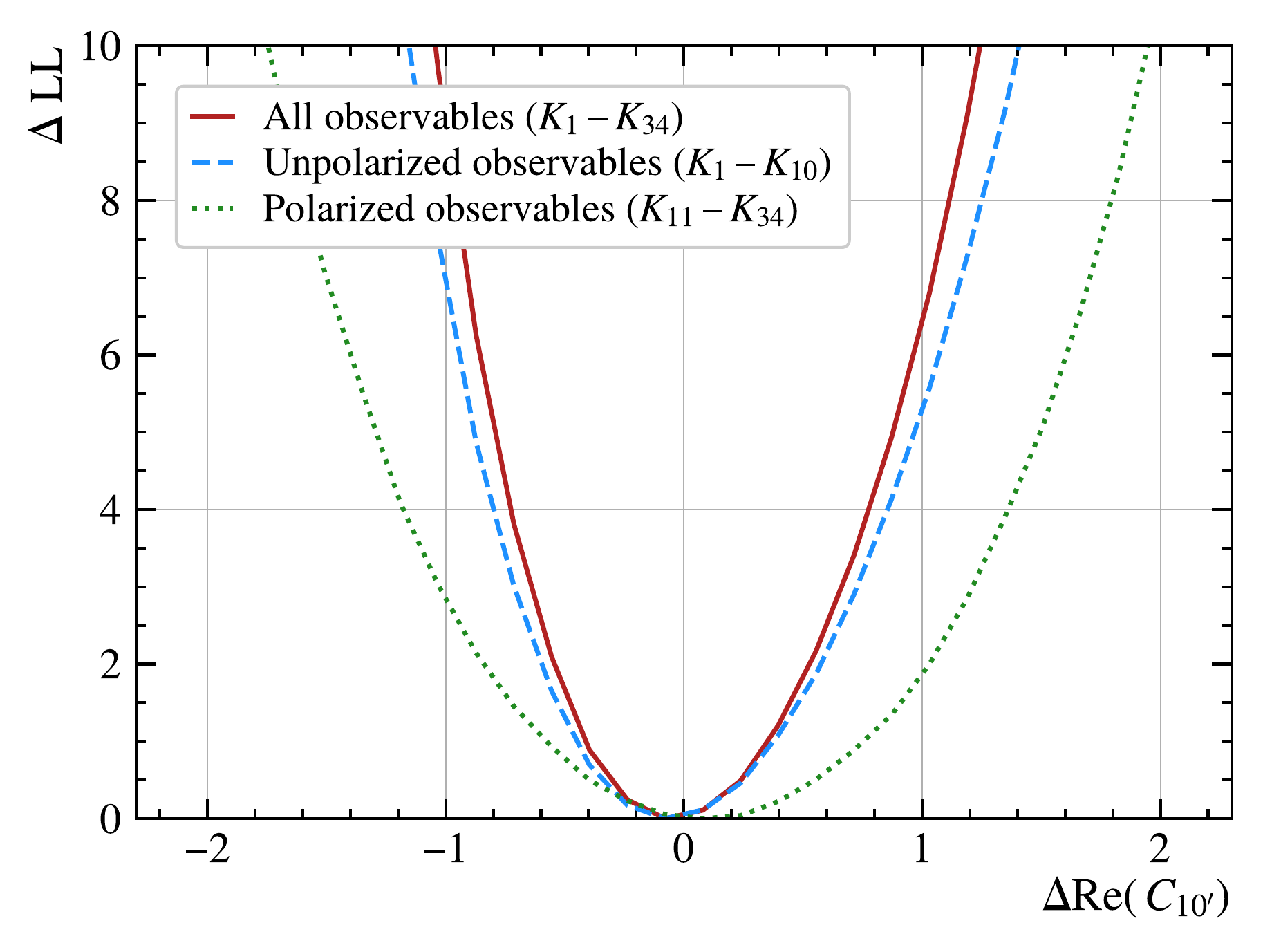}
    \includegraphics[width=0.8\linewidth]{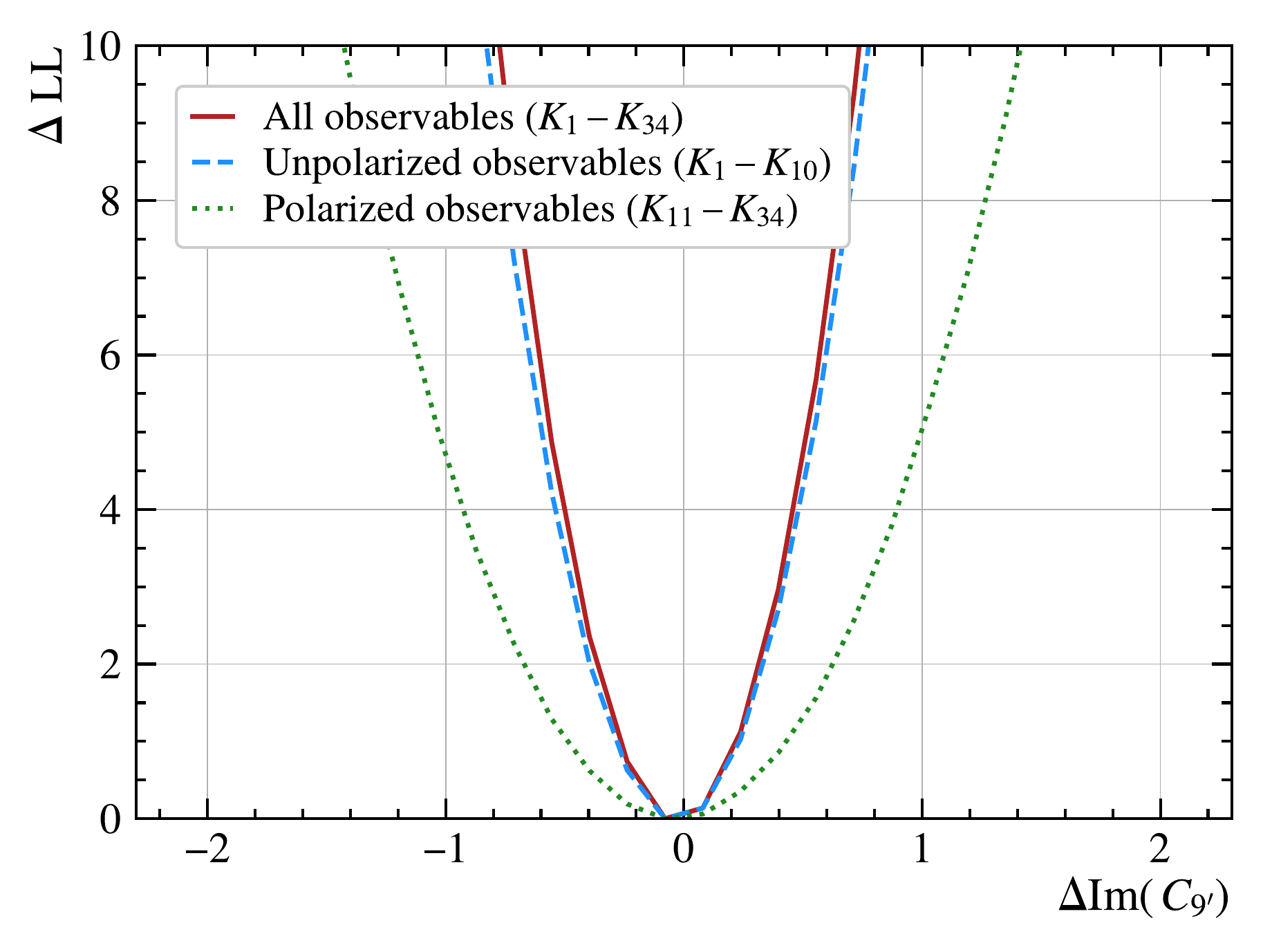}
    \includegraphics[width=0.8\linewidth]{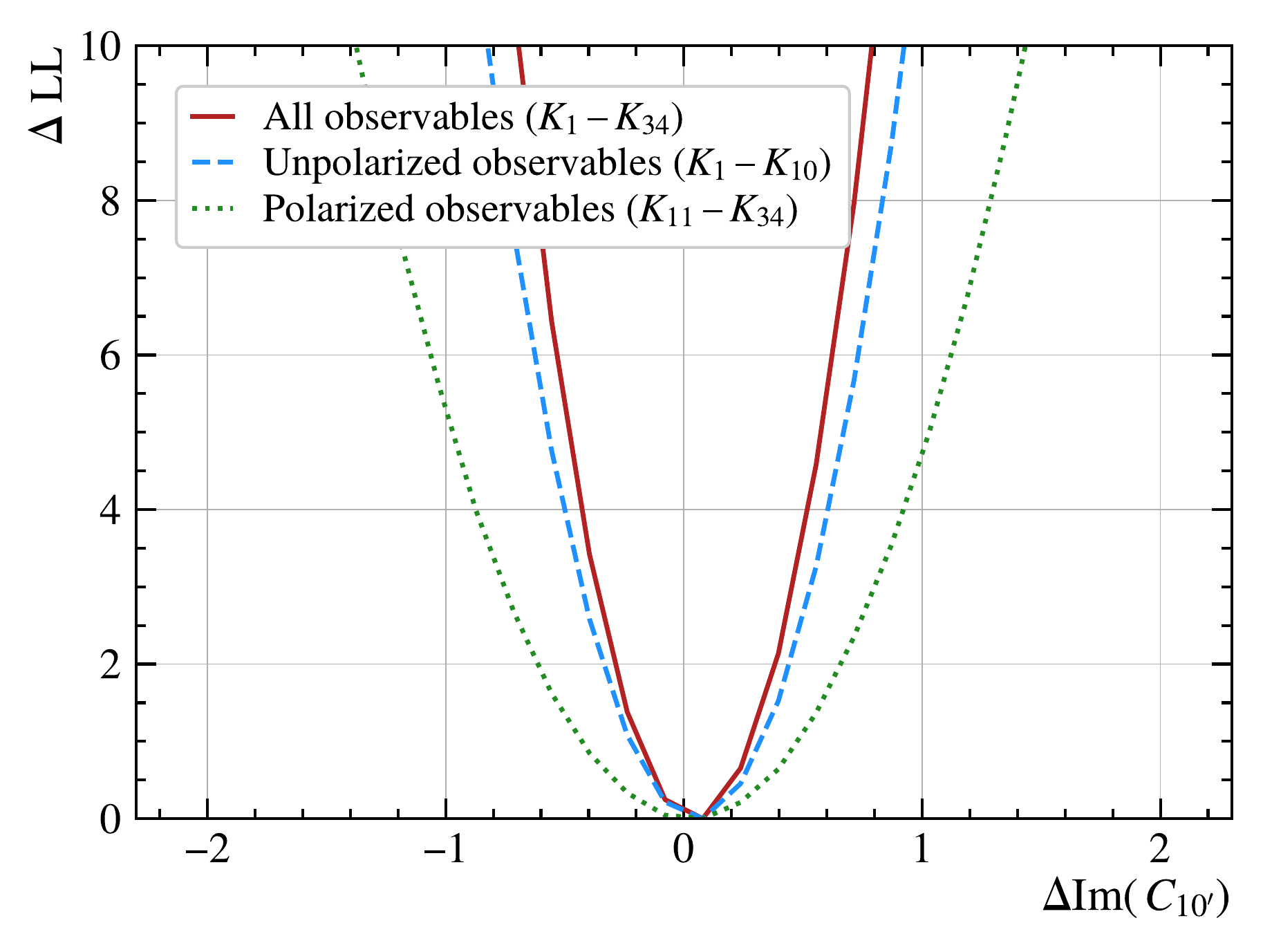}
    \caption{Sensitivity of a fit to the real and imaginary parts of the Wilson coefficients $C_{9^\prime}$ and $C_{10^\prime}$ for a measurement using either all observables, only unpolarized, or only polarized observables assuming perfect particle identification.}
    \label{fig:sensitivity:prime}
\end{figure}

The value of and uncertainty on the Wilson coefficients in the weak effective theory are determined by means of a fit to a set of measurements using the flavor fitting tool \textsc{flavio}~\cite{Straub:2018kue}.
In order to illustrate the sensitivity improvement caused by the polarized observables $K_{11-34}$, the fit is performed three times, once only using the first ten observables, once only using the last 24, and finally to all 34 observables.
In all cases, both analysis bins are considered simultaneously and perfect particle identification is assumed, which represents the uncertainties when the $p-\pi$ separation is better than $2\sigma$.
Figures~\ref{fig:sensitivity:noprime} and \ref{fig:sensitivity:prime} show the resulting scans of the logarithm of the likelihood with respect to the real and imaginary parts of the four most relevant Wilson coefficients $C_{9^{(\prime)}}$ and $C_{10^{(\prime)}}$.
The slight shifts away from the SM at zero are due to a mismatch in the modeling of the local operators in the \textsc{flavio} package with respect to the model used to reweight the simulation.
The disjoint sets of observables dependent and independent of the polarization result in similar sensitivity while their combination improves the fit notably.
It has been checked that the uncertainties on the first ten angular observables do not decrease when fitting a reduced function obtained from integrating over the polarization angles leading to a three-dimensional angular distribution with only ten coefficients.


The fit is also performed for each observable individually with results shown in Appendix~\ref{app:individual} (Figs.~\ref{fig:individual} and \ref{fig:individual:prime}).
Many observables lead to constraints on the real part of $C_9$, this includes both observables with and without dependence on the polarization and in particular $K_3$, $K_6$, $K_{13}$, $K_{24}$, and $K_{28}$.
Only the observable $K_3$ has a strong impact on the real part of $C_{10}$ and the imaginary parts of $C_9$ and $C_{10}$ on its own.
The sensitivity to the real parts of $C_{9^\prime}$ and $C_{10^\prime}$ are strongly dependent on the sensitivity of $K_3$, $K_4$, and $K_8$.
The imaginary parts of $C_{9^\prime}$ and $C_{10^\prime}$ on the other hand are dominated by the sensitivity of $K_{10}$ with notable constraints from $K_4$.
In summary, the likelihood profiles for the unprimed Wilson coefficients shown in Fig~\ref{fig:sensitivity:noprime} are a consequence of considering an ensemble of observables.
The additional information due to considering polarized \Lb baryons leads to a notable improvement of the sensitivity in these cases.
The primed Wilson coefficients on the other hand are very well constrained by $K_3$, $K_4$, and $K_8$.
The addition of aobservables with or without dependence on the polarization leads to only small improvements.

As discussed previously, the projected signal yield for the FCC-ee corresponds to the expected yield by the LHCb experiment sometime during the Upgrade~II run period.
Due to employing a background-free fit instead of a moment analysis with background however, the statistical uncertainties for the FCC-ee measurement correspond more closely to the statistical sensitivity of the angular observables after the Upgrade~II period as estimated in Ref.~\cite{Blake:2017une}.
As a consequence, the sensitivity on the Wilson coefficients shown here for the unpolarized observables serves as a conservative proxy for the expected sensitivity obtained from the final LHCb measurement.

\section{Discussion and conclusion}
This paper presents a study of \LbToppimm decays at the proposed FCC-ee.
The estimated sensitivity on the angular observables is of a  similar magnitude as the sensitivity projected for the LHCb experiment by the end of the LHC lifetime.
Following the access to the full set of polarized angular observables at the FCC-ee however, the sensitivity to the Wilson coefficients is improved as shown in Figs.~\ref{fig:sensitivity:noprime} and \ref{fig:sensitivity:prime}.

The study was performed using DELPHES simulation which is only a proxy for the quantities to be expected in data and simplifies the reconstruction process.
However the real tracking and reconstruction efficiency is expected to be very high given the low-multiplicity $e^+e^-\to Z^0$ events produced at the FCC-ee.
What is more, improvements in tracking and reconstruction methods can be expected to improve in the future for example through the usage of more sophisticated machine learning techniques.

A future study may investigate the potential inclusion of the low dimuon invariant mass region or the region between the \jpsi and \psitwos resonances as well as finer binning of the region between the $\phi$ and \jpsi resonances.
Moreover, considering larger $Z^0\to\qqbar$ simulation samples may be useful for higher confidence in the negligence of combinatorial background.
Similarly, a simple refit of the decay chain while fixing the dihadron invariant mass and the four-body invariant mass to the known \Lz and \Lb masses might enable further suppression of such backgrounds.
New statistical analysis methods may improve the achievable precision of the angular coefficients and developments in the understanding of form factors may improve the accessibility of the Wilson coefficients on data directly.
Finally, future studies may dedicate efforts to aligning the models used to obtain the theory predictions and experimental simulation to mitigate the shifts in the Wilson coefficients.

While the exact numbers presented in this study will change in future iterations, it is clear that the access to all 34 angular observables due to the nonzero polarization of \Lb baryons produced in $Z^0$ decays provides a significant increase to the sensitivity to the Wilson coefficients.

\section*{A\lowercase{cknowledgements}}
We would like to express our gratitude to Sebastian Schmitt for providing a parallelized fitting setup for the \textsc{flavio} package.
We also want to thank our colleagues in the FCC organization for providing the simulation samples, software, and advice, in particular Jan Eysermans, Aidan Wiederhold, and Xunwu Zuo.
We would like to acknowledge the US National Science Foundation, whose funding under Award No. 2310073 has helped support this work.

\section*{Data Availability}
The data that support the findings of this article are not publicly available.
The data are available from the authors upon reasonable request.

\clearpage
\input{appendix}

\bibliography{apssamp}

\end{document}

%% file: symbols-def.tex
\def\PLambda     {\ensuremath{\Lambda}\xspace}
\def\PB          {\ensuremath{B}\xspace}
\def\PK          {\ensuremath{K}\xspace}
\def\PJ          {\ensuremath{J}\xspace}
\def\Pb          {\ensuremath{b}\xspace}
\def\Ps          {\ensuremath{s}\xspace}
\def\Pc          {\ensuremath{c}\xspace}

\def\Pq          {\ensuremath{q}\xspace}
\def\Pp          {\ensuremath{p}\xspace}
\def\Pmu         {\ensuremath{\mu}\xspace}
\def\Ppi         {\ensuremath{\pi}\xspace}

\def\Ppsi        {\ensuremath{\psi}\xspace}
\def\Pphi        {\ensuremath{\phi}\xspace}

\def\mup        {{\ensuremath{\Pmu^+}}\xspace}
\def\mun        {{\ensuremath{\Pmu^-}}\xspace}

\def\mumu       {{\ensuremath{\Pmu^+\Pmu^-}}\xspace}

\def\ellm       {{\ensuremath{\ell^-}}\xspace}
\def\ellp       {{\ensuremath{\ell^+}}\xspace}

\def\ellell     {\ensuremath{\ell^+ \ell^-}\xspace}

\def\quark     {{\ensuremath{\Pq}}\xspace}
\def\quarkbar  {{\ensuremath{\overline \quark}}\xspace}
\def\qqbar     {{\ensuremath{\quark\quarkbar}}\xspace}

\def\squark    {{\ensuremath{\Ps}}\xspace}

\def\cquark    {{\ensuremath{\Pc}}\xspace}
\def\cquarkbar {{\ensuremath{\overline \cquark}}\xspace}
\def\ccbar     {{\ensuremath{\cquark\cquarkbar}}\xspace}
\def\bquark    {{\ensuremath{\Pb}}\xspace}
\def\bquarkbar {{\ensuremath{\overline \bquark}}\xspace}
\def\bbbar     {{\ensuremath{\bquark\bquarkbar}}\xspace}

\def\pion   {{\ensuremath{\Ppi}}\xspace}

\def\pip    {{\ensuremath{\pion^+}}\xspace}
\def\pim    {{\ensuremath{\pion^-}}\xspace}

\def\kaon    {{\ensuremath{\PK}}\xspace}
\def\Kbar    {{\ensuremath{\overline{\PK}}}\xspace}

\def\KorKbar {\kern \thebaroffset\optbar{\kern -\thebaroffset \PK}{}\xspace}

\def\Kp      {{\ensuremath{\kaon^+}}\xspace}
\def\Km      {{\ensuremath{\kaon^-}}\xspace}

\def\KS      {{\ensuremath{\kaon^0_{\mathrm{S}}}}\xspace}

\def\Kstarz  {{\ensuremath{\kaon^{*0}}}\xspace}
\def\Kstarzb {{\ensuremath{\Kbar{}^{*0}}}\xspace}

\def\Kstarp  {{\ensuremath{\kaon^{*+}}}\xspace}

\newcommand{\phiz}{\ensuremath{\Pphi}\xspace}

\def\B       {{\ensuremath{\PB}}\xspace}
\def\Bbar    {{\ensuremath{\overline{\PB}}}\xspace}

\def\BorBbar {\kern \thebaroffset\optbar{\kern -\thebaroffset \PB}\xspace}
\def\Bz      {{\ensuremath{\B^0}}\xspace}

\def\Bd      {{\ensuremath{\B^0}}\xspace}
\def\Bdb     {{\ensuremath{\Bbar{}^0}}\xspace}
\def\BdorBdbar {\kern \thebaroffset\optbar{\kern -\thebaroffset \Bd}\xspace}
\def\Bu      {{\ensuremath{\B^+}}\xspace}

\def\Bp      {{\ensuremath{\Bu}}\xspace}

\def\Bs      {{\ensuremath{\B^0_\squark}}\xspace}

\def\BsorBsbar {\kern \thebaroffset\optbar{\kern -\thebaroffset \Bs}\xspace}

\def\jpsi     {{\ensuremath{{\PJ\mskip -3mu/\mskip -2mu\Ppsi}}}\xspace}
\def\psitwos  {{\ensuremath{\Ppsi{(2S)}}}\xspace}

\def\proton      {{\ensuremath{\Pp}}\xspace}

\def\Lz          {{\ensuremath{\PLambda}}\xspace}
\def\Lbar        {{\ensuremath{\overline{\PLambda}}}\xspace}
\def\Lb           {{\ensuremath{\Lz^0_\bquark}}\xspace}
\def\Lbbar        {{\ensuremath{\Lbar{}^0_\bquark}}\xspace}

\def\CP                {{\ensuremath{C\!P}}\xspace}

\newcommand{\decay}[2]{\mbox{\ensuremath{#1\!\to #2}}\xspace}

\def\to                 {\ensuremath{\rightarrow}\xspace}

\def\LbToppimm    {\decay{\Lb}{\Lz(\to p\pim)\mup\mun}}
\def\BdTopipimm    {\decay{\Bd}{\KS(\to \pip\pim)\mup\mun}}

\def\BdbToKstmm   {\decay{\Bdb}{\Kstarzb\mup\mun}}
\def\bsll     {\decay{\bquark}{\squark \ell^+ \ell^-}}

\newcommand{\aunit}[1]{\ensuremath{\text{\,#1}}}      

\newcommand{\tev}{\aunit{Te\kern -0.1em V}\xspace}
\newcommand{\gev}{\aunit{Ge\kern -0.1em V}\xspace}
\newcommand{\mev}{\aunit{Me\kern -0.1em V}\xspace}
\newcommand{\kev}{\aunit{ke\kern -0.1em V}\xspace}
\newcommand{\ev}{\aunit{e\kern -0.1em V}\xspace}

\newcommand{\mevc}{\ensuremath{\aunit{Me\kern -0.1em V\!/}c}\xspace}
\newcommand{\gevc}{\ensuremath{\aunit{Ge\kern -0.1em V\!/}c}\xspace}
\newcommand{\mevcc}{\ensuremath{\aunit{Me\kern -0.1em V\!/}c^2}\xspace}
\newcommand{\gevcc}{\ensuremath{\aunit{Ge\kern -0.1em V\!/}c^2}\xspace}

\def\mm   {\aunit{mm}\xspace}

\def\th{{\ensuremath{\theta_h}}\xspace}
\def\tm{{\ensuremath{\theta_\mu}}\xspace}
\def\tp{{\ensuremath{\theta_\parallel}}\xspace}
\def\phih{{\ensuremath{\phi_h}}\xspace}
\def\phim{{\ensuremath{\phi_\mu}}\xspace}
\def\qprime{{\ensuremath{q^\prime}}\xspace}
\def\phsp{{\ensuremath{\Phi}}\xspace}
\def\qsq{\ensuremath{q^2}\xspace}

\def\deriv {\ensuremath{\mathrm{d}}}

%% file: contaminations.tex
\begin{tabular}{c|l|lll|l}
\toprule
Process & Production & Acceptance & Topology & Total & Selected \\
\midrule
$Z^0\to b\bar{b}$ & $9.00\cdot10^{11}$ & $3.0\cdot10^{-4}$ & $2.6\cdot10^{-5}$ & $<{4.2}\cdot10^{-9}$ & $<3759$ \\
$Z^0\to c\bar{c}$ & $7.20\cdot10^{11}$ & $3.2\cdot10^{-5}$ & $2.0\cdot10^{-6}$ & $<{4.8}\cdot10^{-9}$ & $<3450$ \\
$Z^0\to s\bar{s}$ & $9.00\cdot10^{11}$ & $4.2\cdot10^{-6}$ & $7.6\cdot10^{-8}$ & $<{3.7}\cdot10^{-9}$ & $<3300$ \\
$Z^0\to u\bar{u}/d\bar{d}$ & $1.62\cdot10^{12}$ & $4.4\cdot10^{-6}$ & $8.0\cdot10^{-8}$ & $<{5.4}\cdot10^{-9}$ & $<8799$ \\
\midrule
\BdTopipimm & $1.68\cdot10^{5}$ & $1.34$ & $0.68$ & $2.6\cdot10^{-2}$ & 4438 \\
Background with $\LbToppimm$ & $9.31\cdot10^{4}$ & $0.49$ & $3.2\cdot10^{-3}$ & $6.7\cdot10^{-5}$ & 6 \\
\midrule
$\LbToppimm$ & $9.31\cdot10^{4}$ & $0.81$ & $0.63$ & $0.42$ & 38895 \\
\bottomrule\end{tabular}

%% file: appendix.tex
\onecolumngrid
\appendix
\section{\uppercase{Reweighted distributions}}\label{app:distributions}
\begin{center}
\begin{figure}[h]
    \centering
    \includegraphics[width=0.4\textwidth]{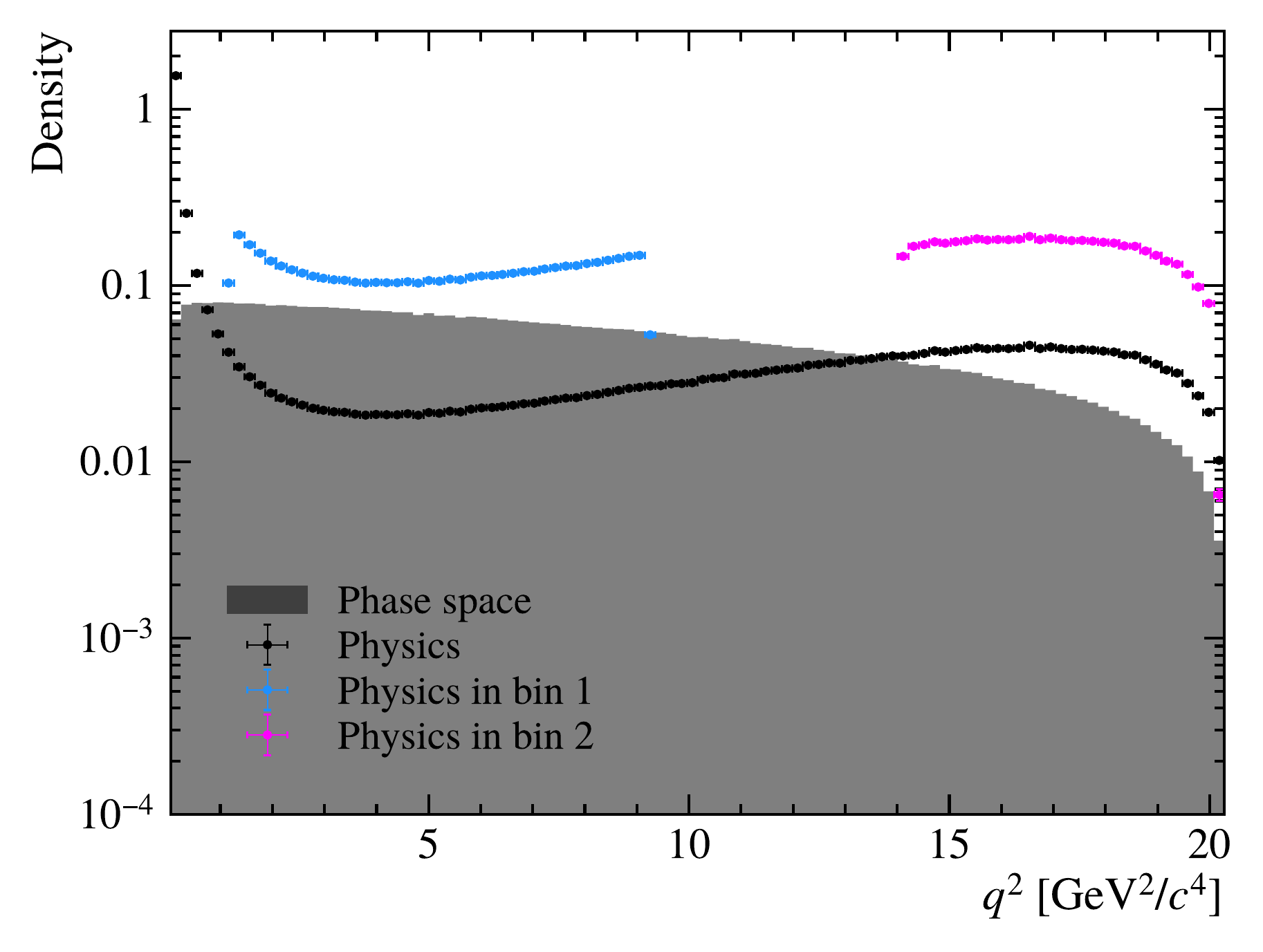}%
    \includegraphics[width=0.4\textwidth]{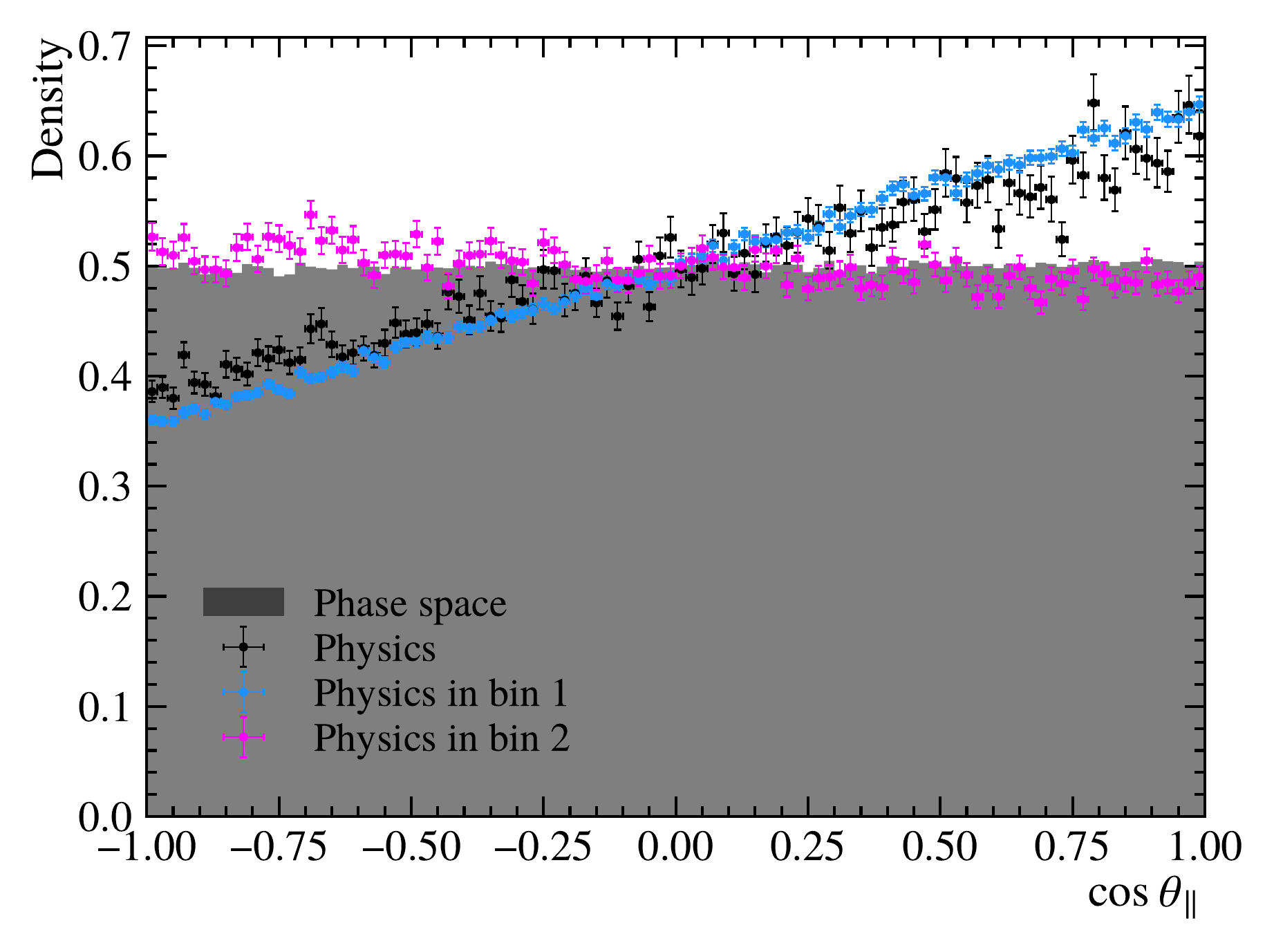}
    \includegraphics[width=0.4\textwidth]{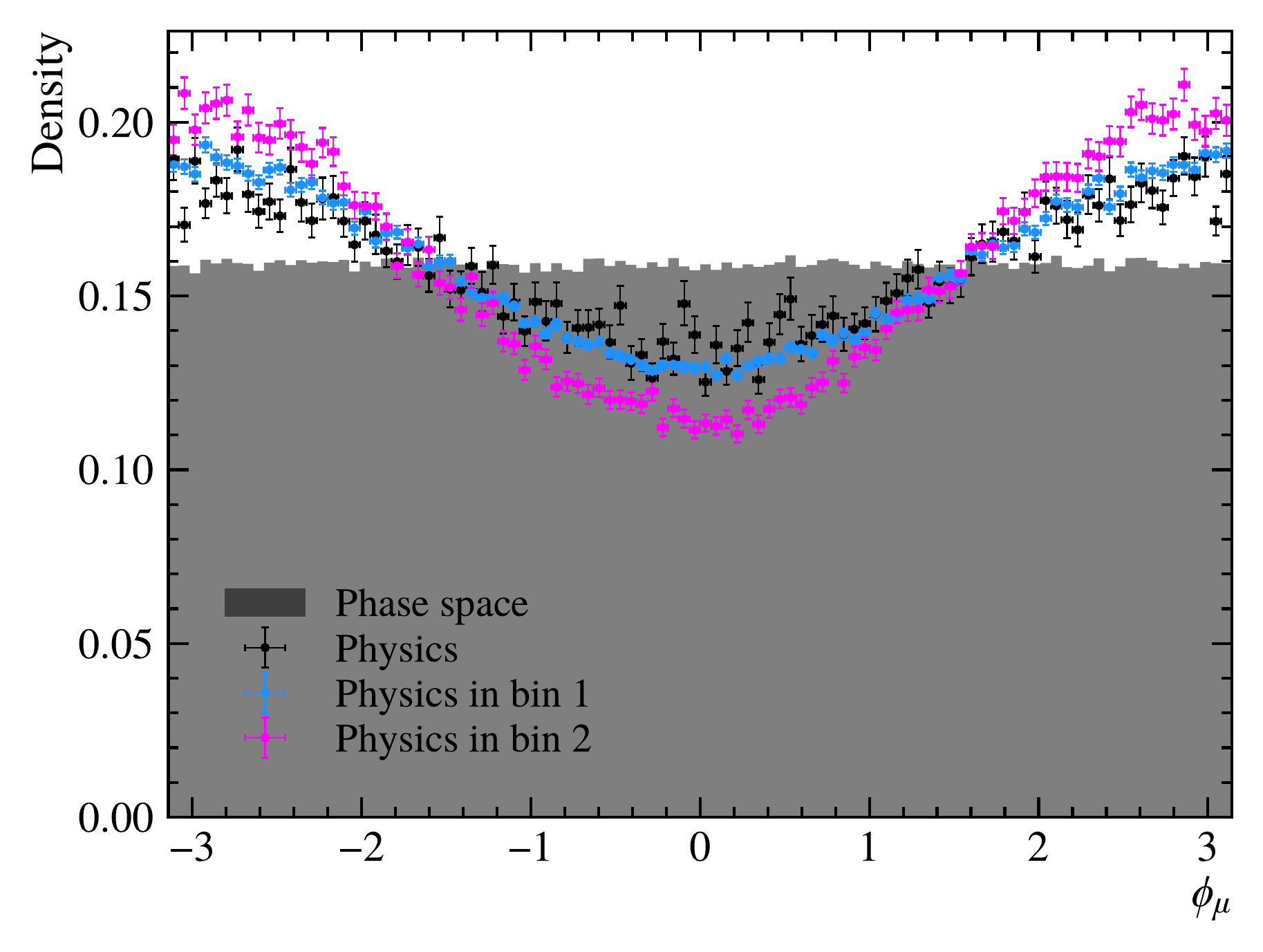}%
    \includegraphics[width=0.4\textwidth]{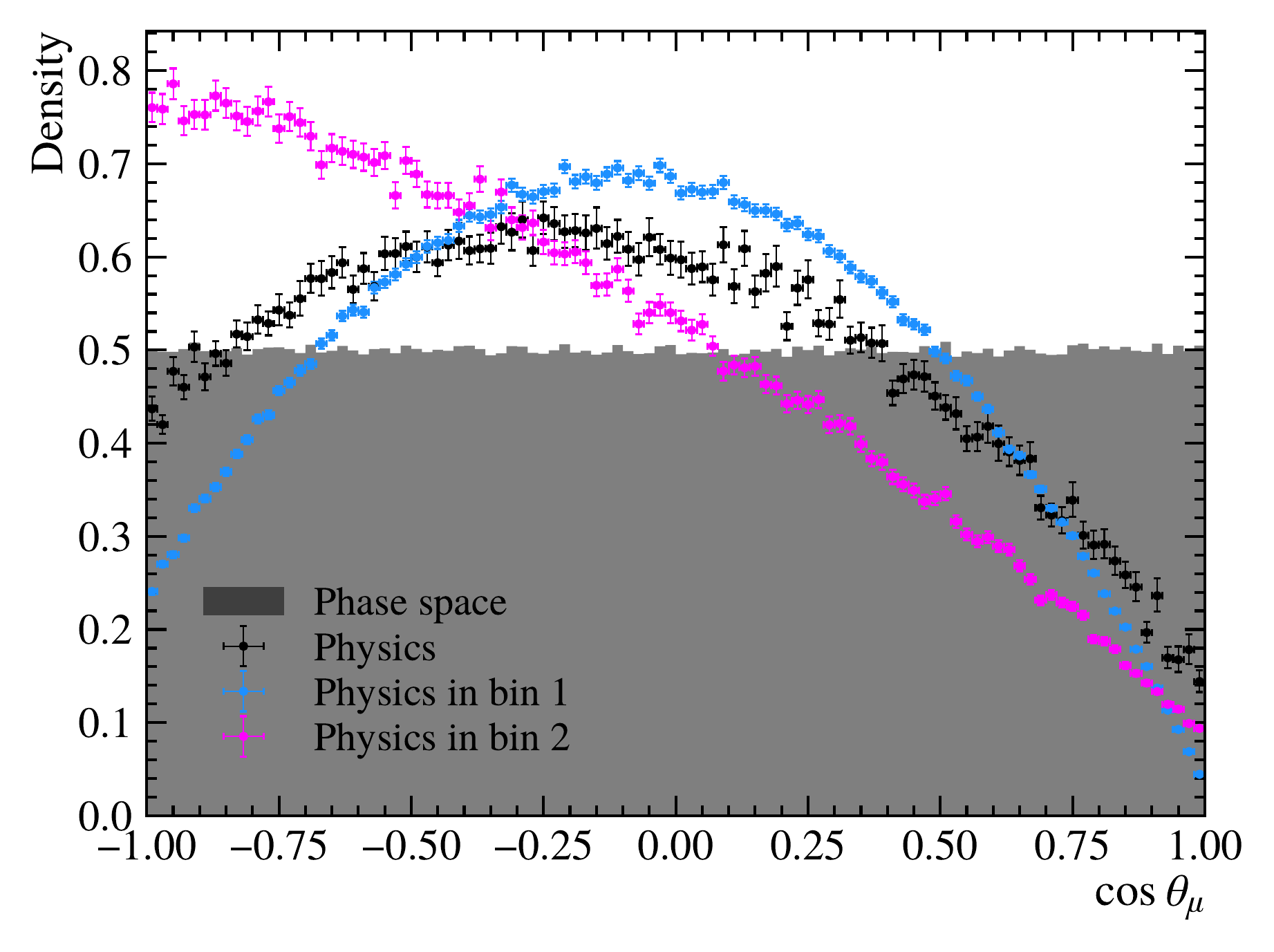}
    \includegraphics[width=0.4\textwidth]{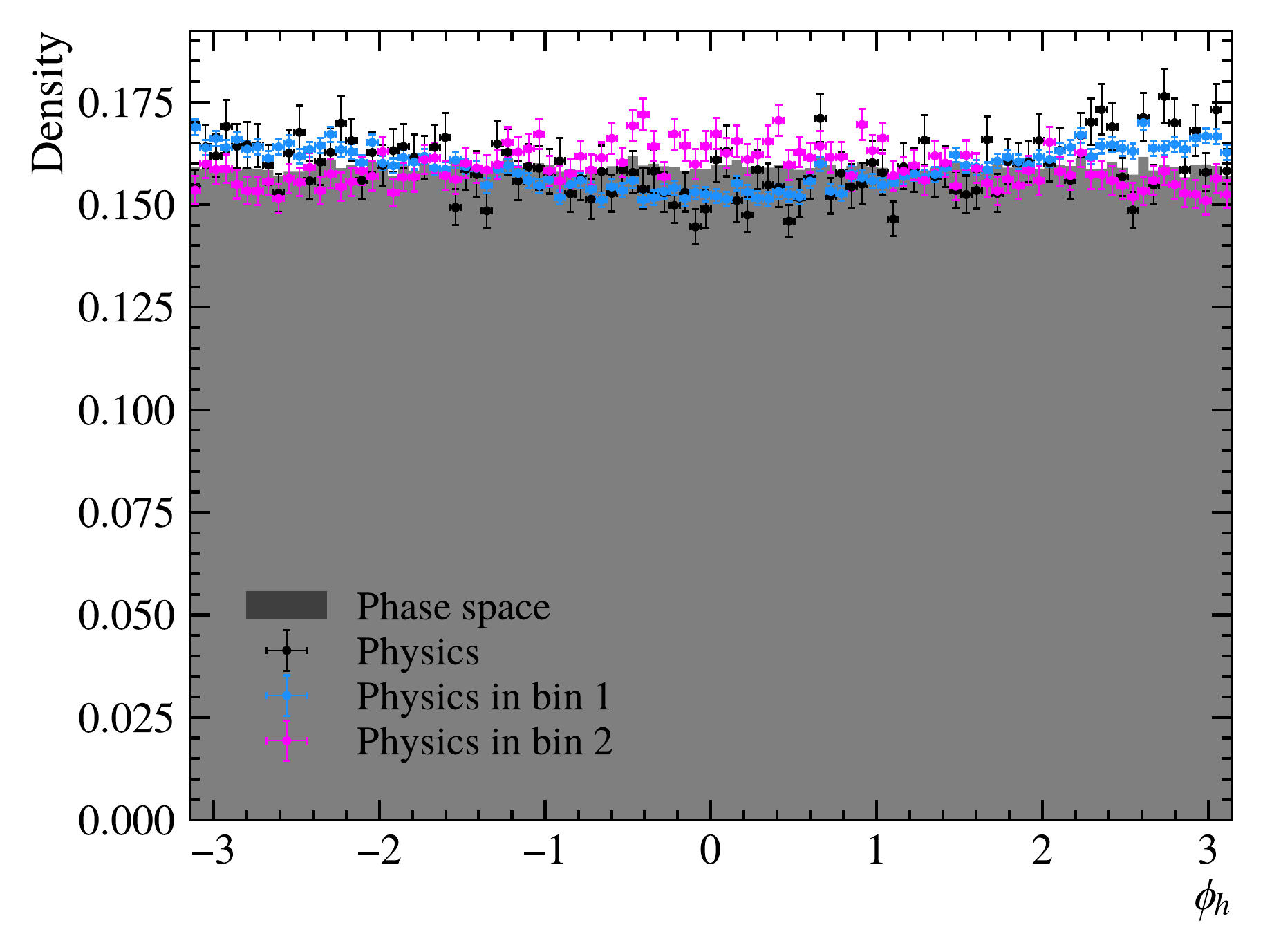}%
    \includegraphics[width=0.4\textwidth]{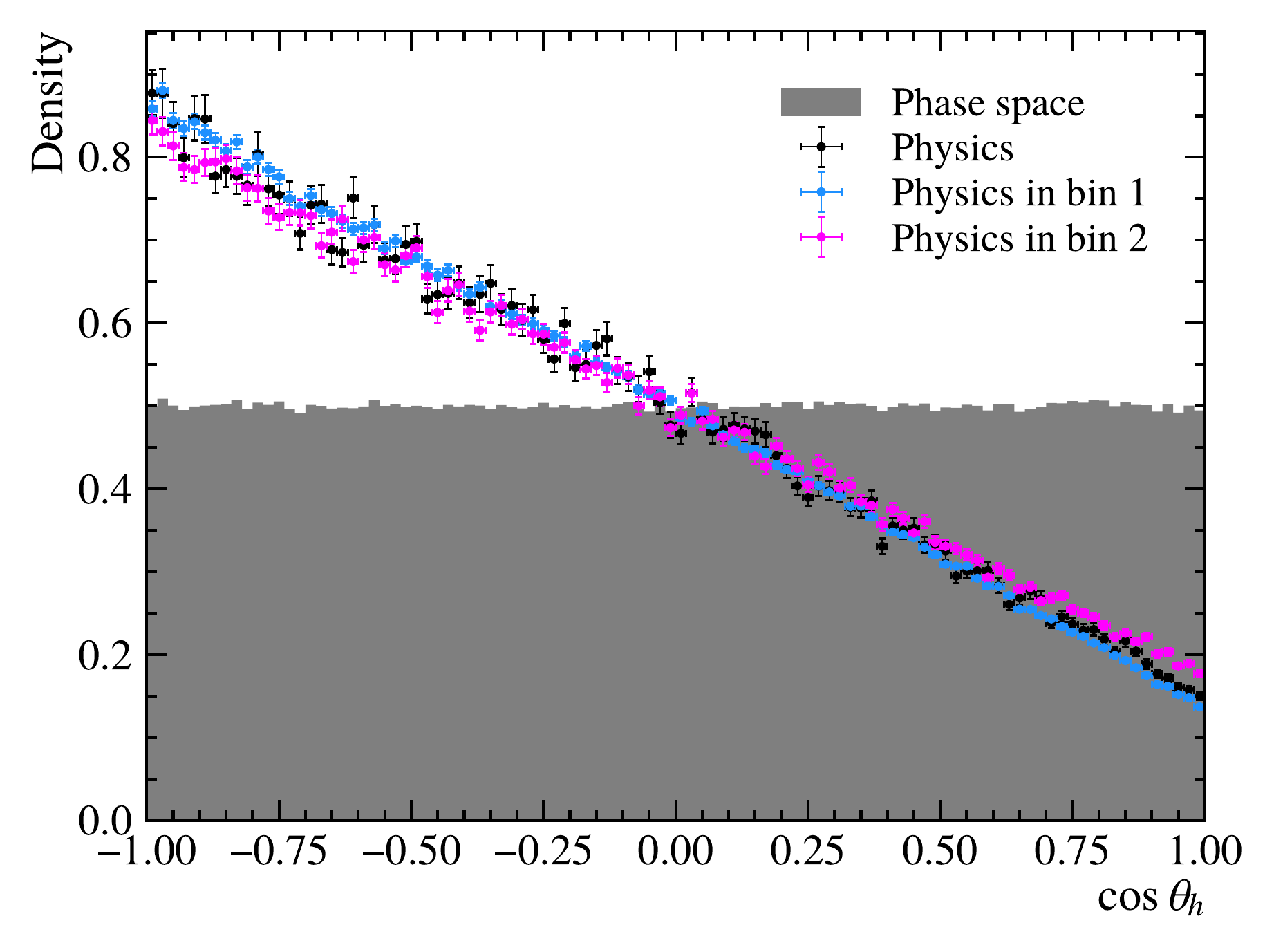}
    \caption{Distribution of the dimuon invariant-mass squared, \qsq, and the five angles in the generated sample (gray) as well as weighted by the Standard Model distribution across the full phase space (black) and in the two analysis bins (blue and magenta).}
    \label{fig:distributions}
\end{figure}
\end{center}
\clearpage
\section{\uppercase{Invariant mass resolutions}}\label{app:resolution}
\begin{figure}[h]
    \centering
    \includegraphics[width=0.4\textwidth]{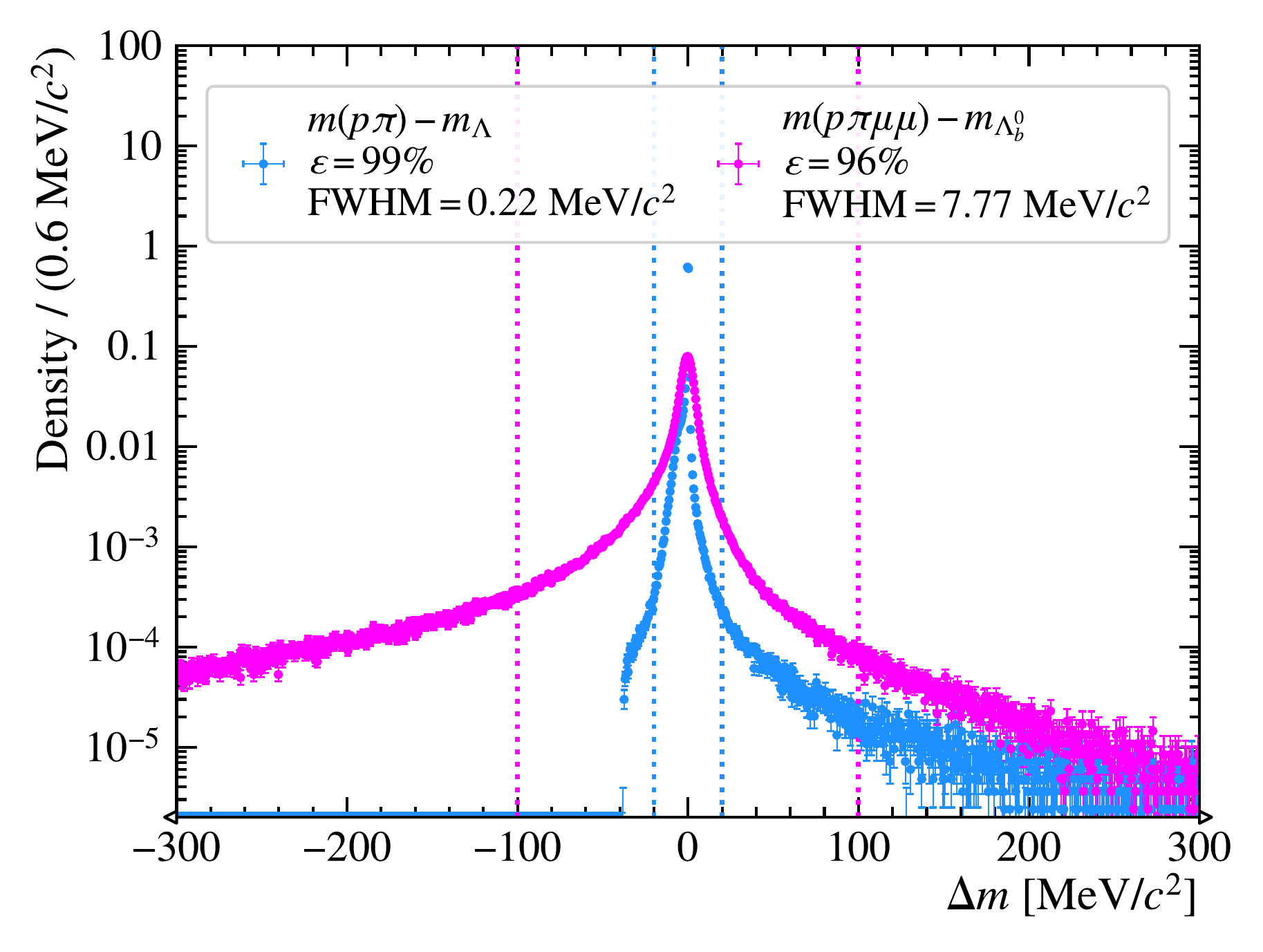}%
    \includegraphics[width=0.4\textwidth]{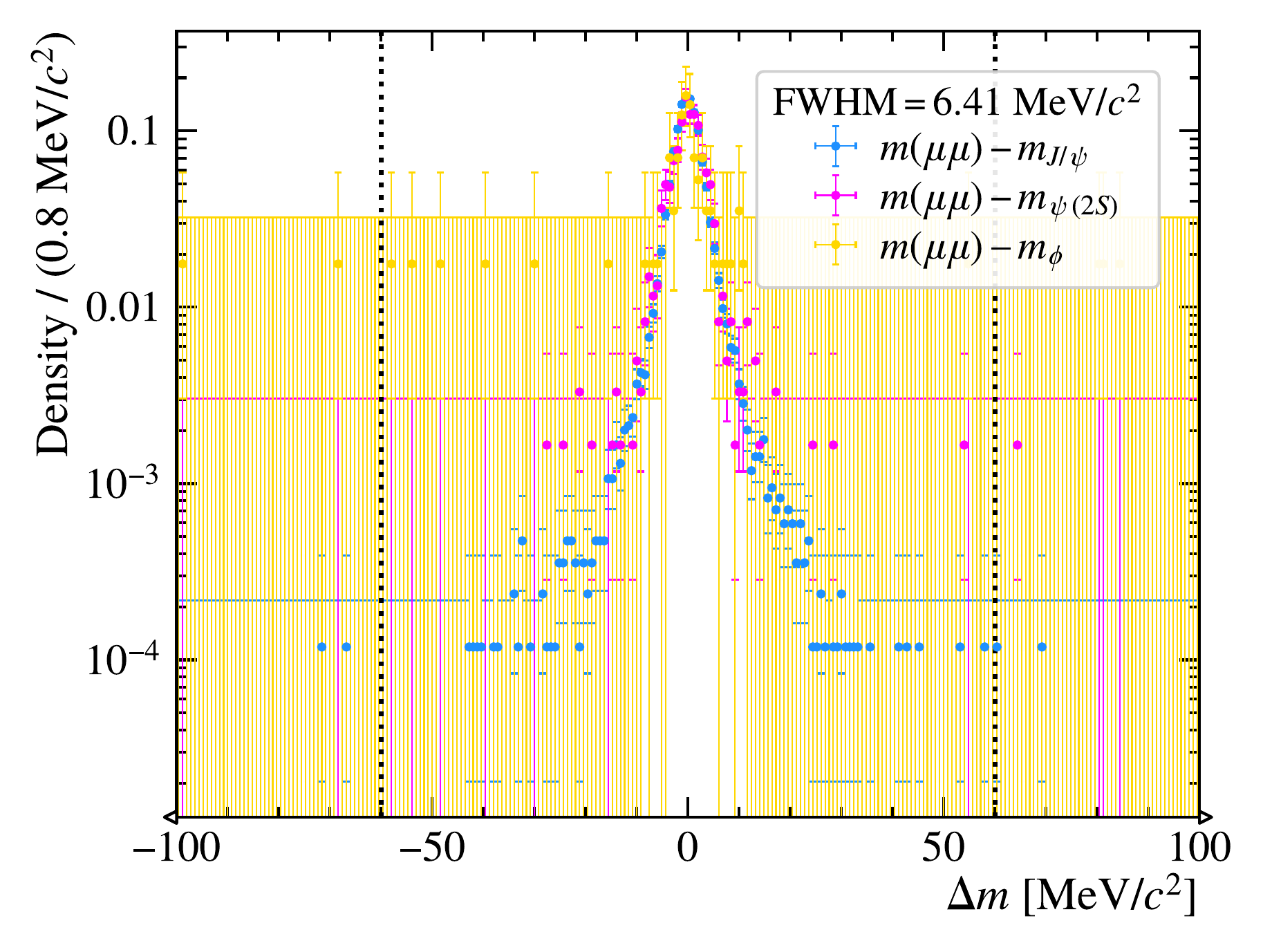}
    \caption{Difference of the reconstructed and true mass of the baryons (left) and dimuon resonances (right).
    The windows used for the selection of the signal (left) and the suppression of the \qqbar resonances (right) are shown as dotted lines.
    The legends include the efficiency of selecting only the region between the lines ($\varepsilon$) as well as the full width at half maximum (FWHM) of the peaks as an indication for the resolution.}
    \label{fig:width}
\end{figure}

\clearpage
\section{\uppercase{Numerical results for the angular coefficients}}\label{app:observables}
\begin{table}[h]
    \centering
    \caption{Summary of the estimated sensitivity on the angular observables $K_i$ assuming the Standard Model in bins (left) 1 and (right) 2.
    The three rightmost columns represent the total uncertainty at a level of $p-\pi$ separation of $\kappa\sigma$.
    A particle separation of at least $2\sigma$ results in values compatible with perfect particle identification ($\infty\sigma$).}
    \label{tab:observables}
    \begin{minipage}{.47\textwidth}
    \input{sensitivity_K_bin1}
    \end{minipage}%
    \begin{minipage}{.47\textwidth}
    \input{sensitivity_K_bin2}
    \end{minipage}
\end{table}
\clearpage

\section{\uppercase{Wilson coefficient fits to individual observables}}\label{app:individual}
\begin{figure}[h]
    \centering
    \includegraphics[width=0.4\textwidth]{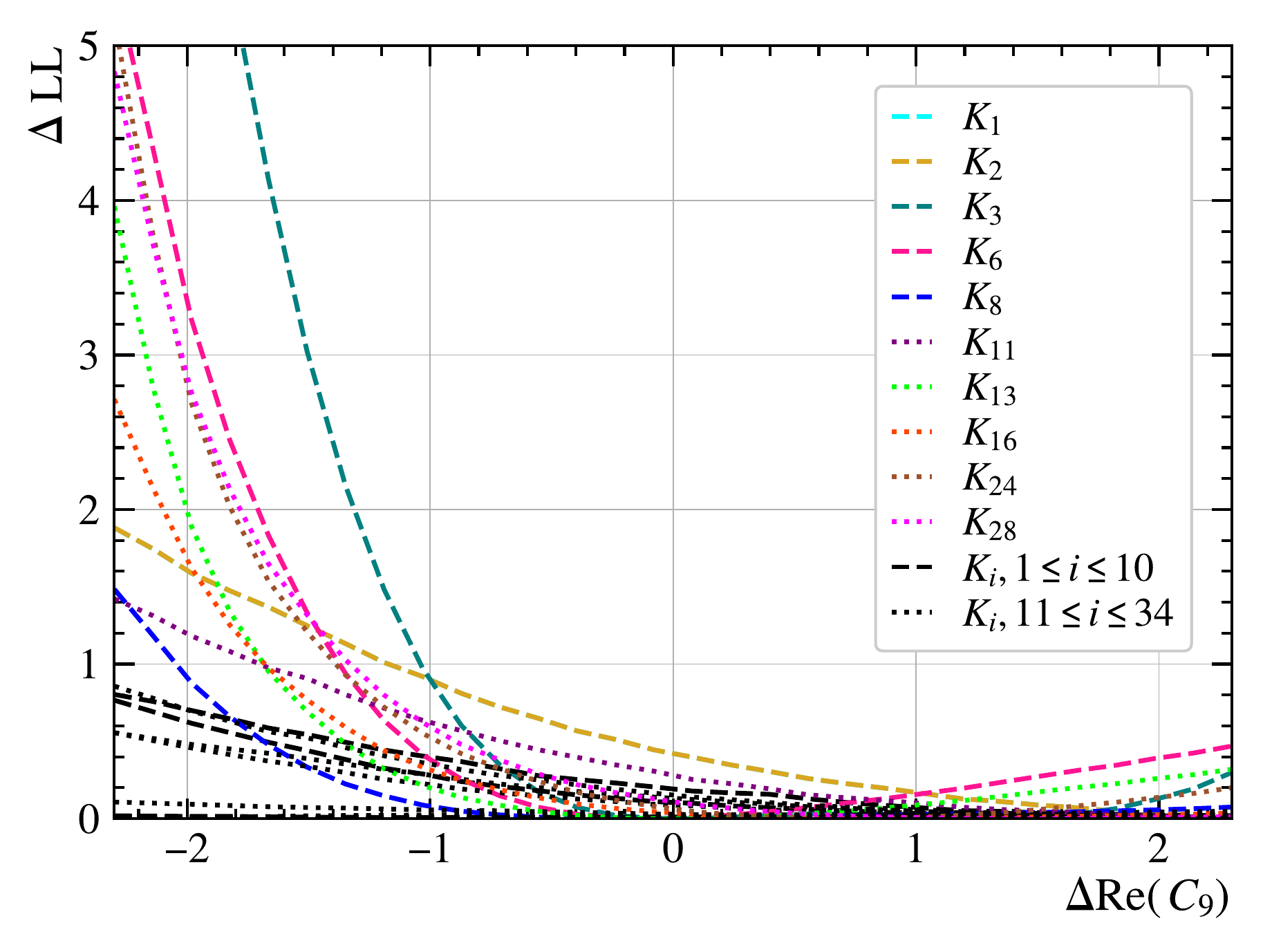}~
    \includegraphics[width=0.4\textwidth]{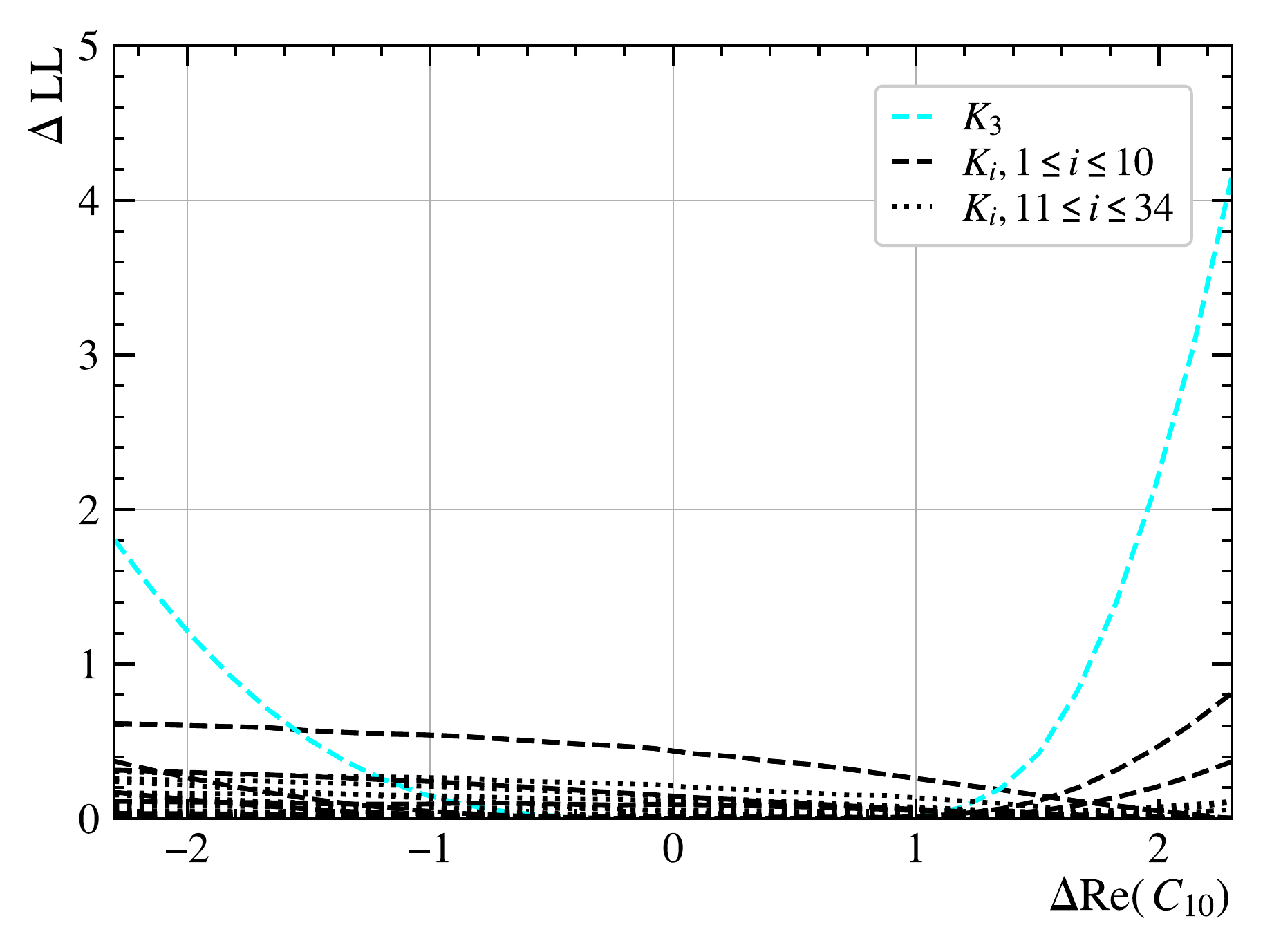}
    \includegraphics[width=0.4\textwidth]{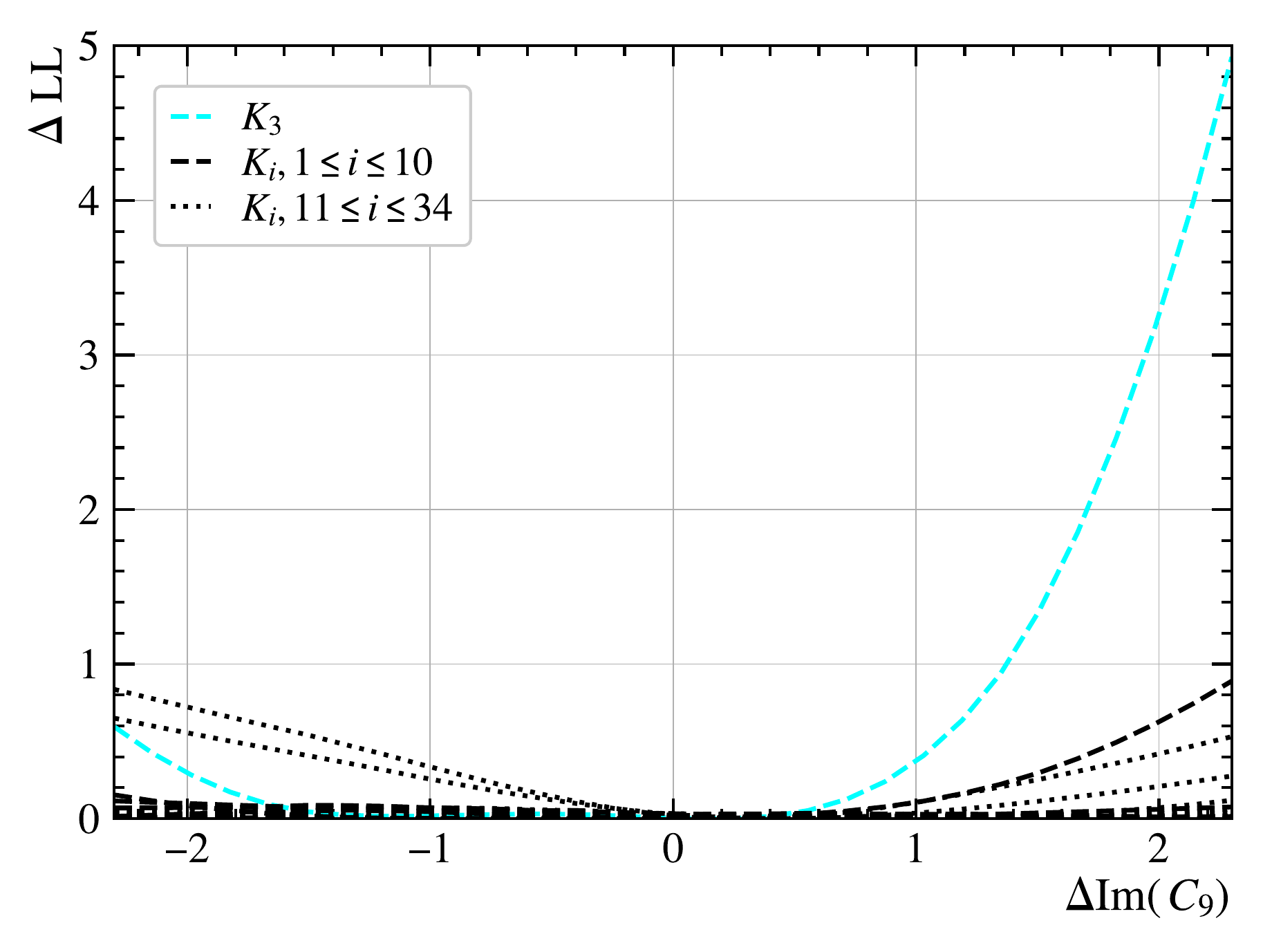}~
    \includegraphics[width=0.4\textwidth]{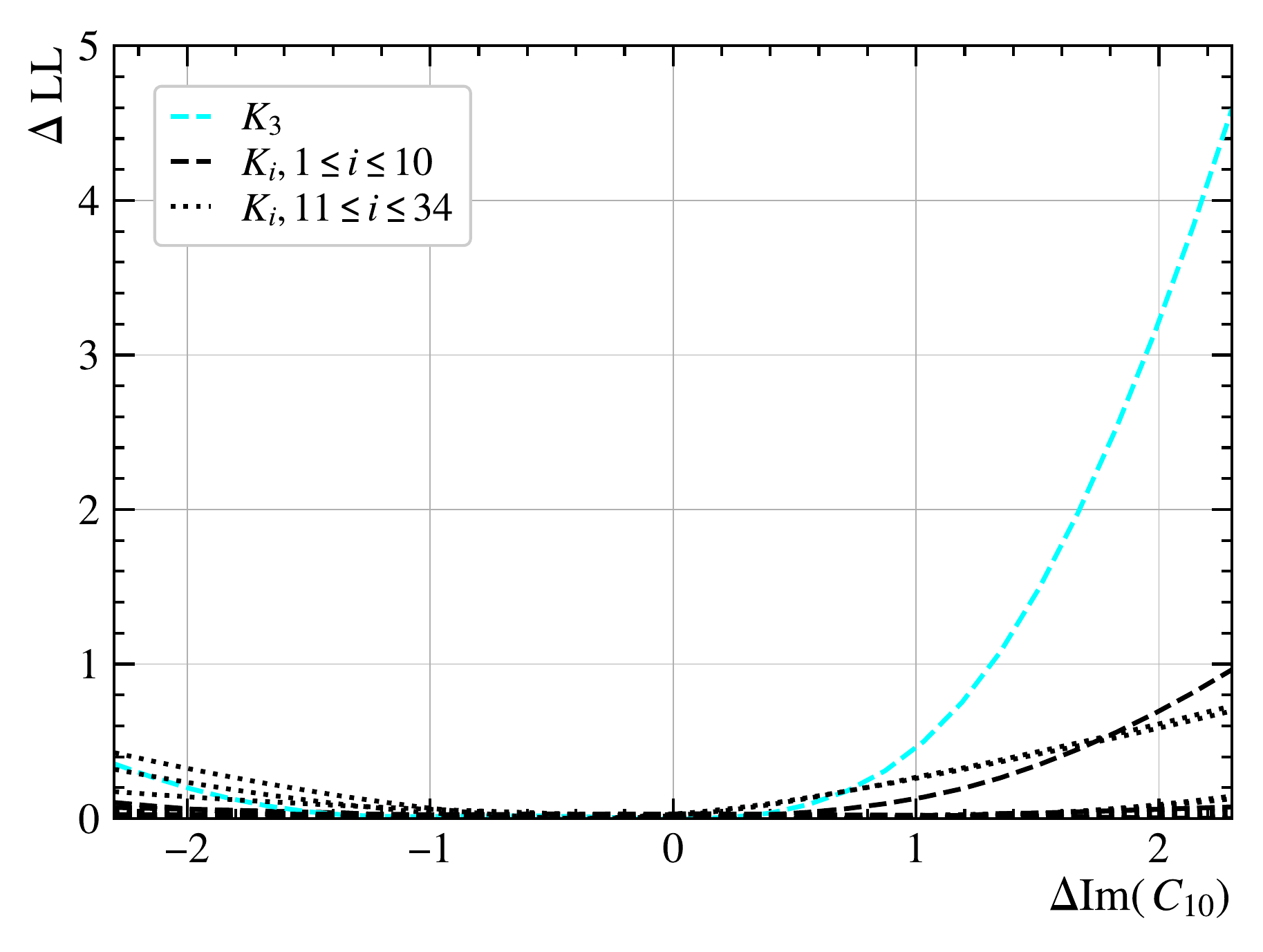}
    \caption{Sensitivity of a fit to the real and imaginary parts of the Wilson coefficients ${C}_9$ and ${C}_{10}$ to each observable individually. Observables for which the logarithmic likelihood difference is less than one across the shown range are plotted as black lines. Dashed (dotted) lines correspond to observables that are independent of (depend on) the polarization. The profiles for $K_1$ and $K_2$ are identical as consequence of the normalization $\tfrac{\deriv\Gamma}{\deriv\qsq} = 2K_1+K_2$.}
    \label{fig:individual}
\end{figure}
\begin{figure}[h]
    \centering
    \includegraphics[width=0.4\textwidth]{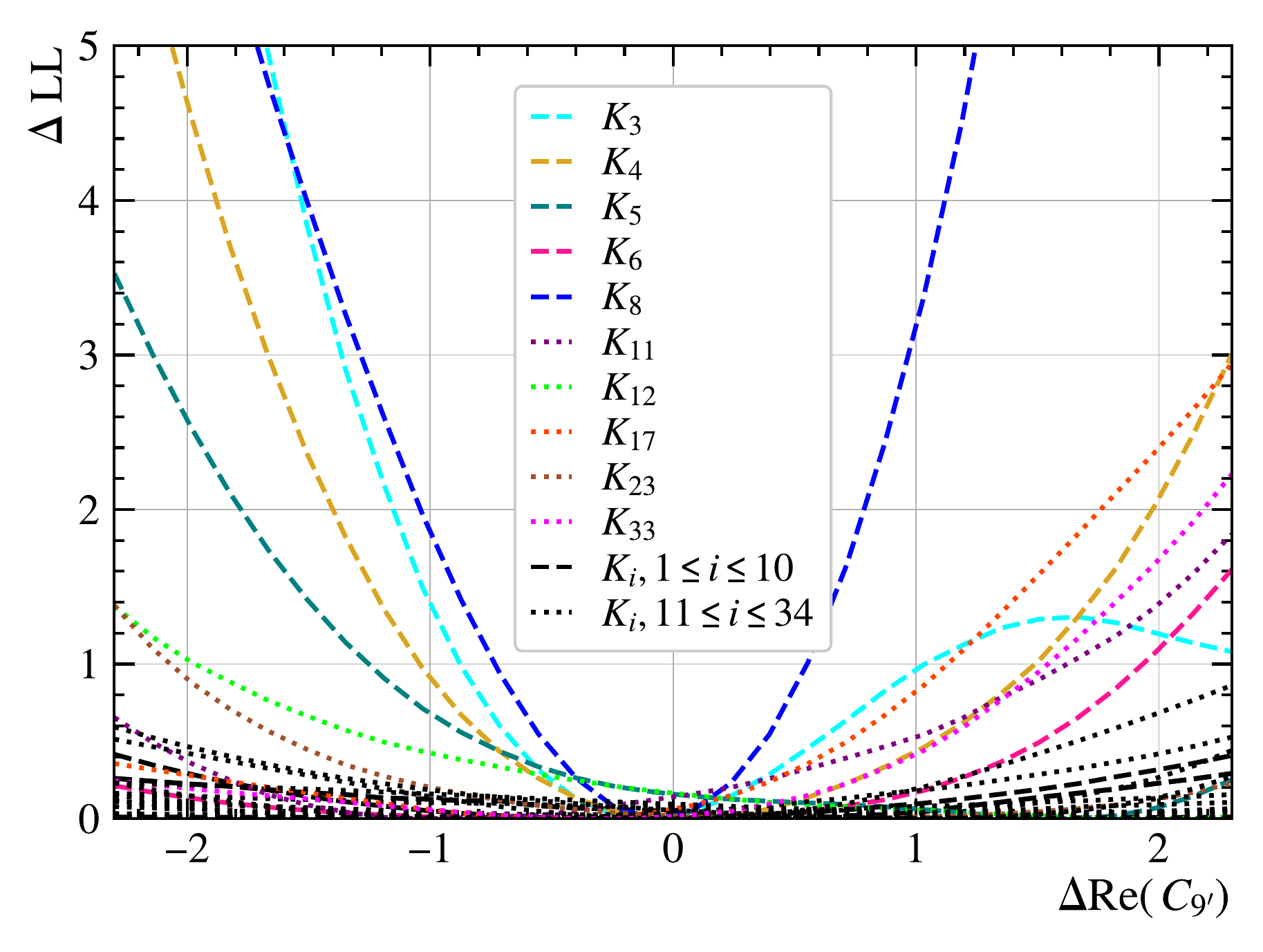}~
    \includegraphics[width=0.4\textwidth]{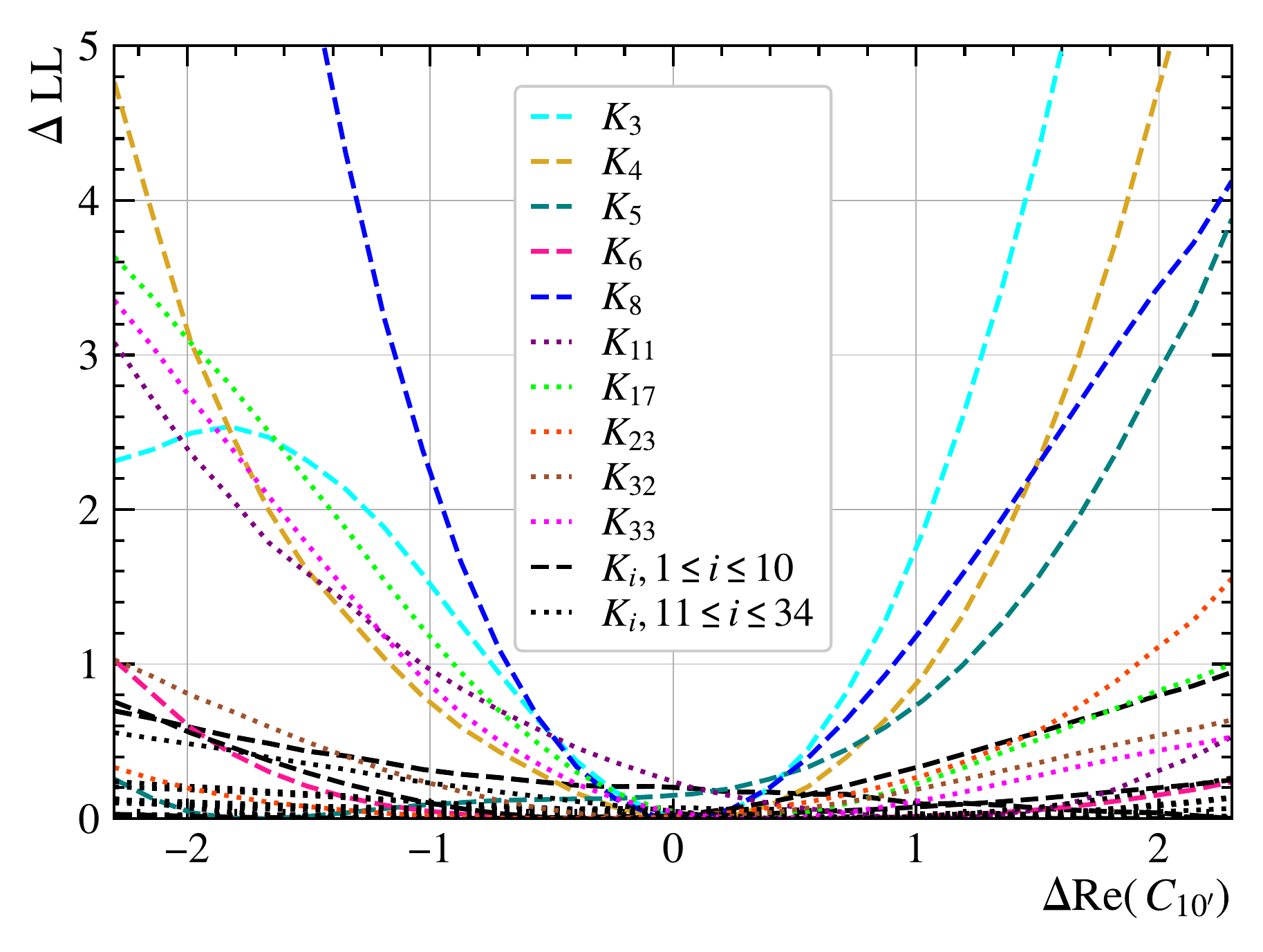}
    \includegraphics[width=0.4\textwidth]{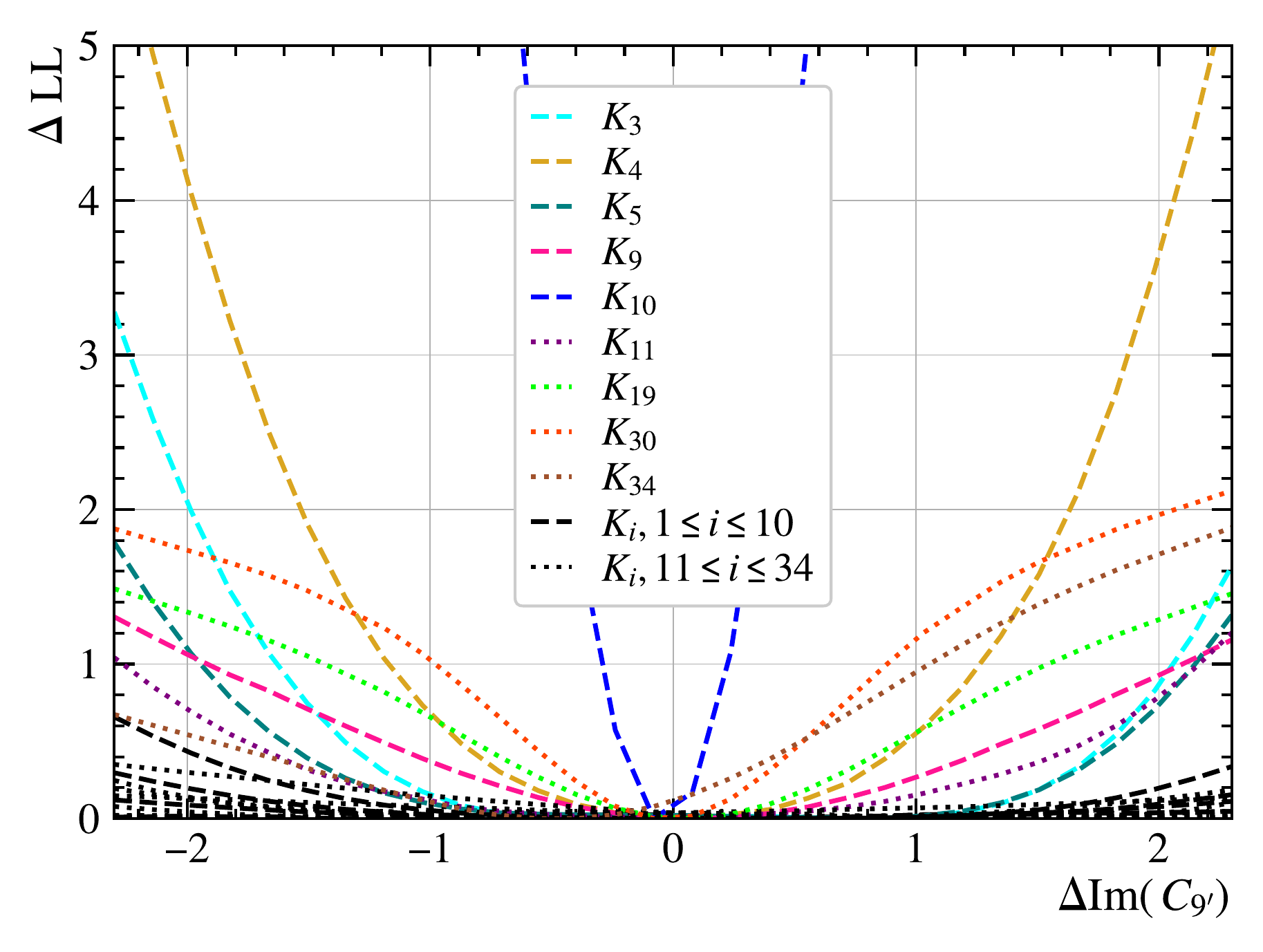}~
    \includegraphics[width=0.4\textwidth]{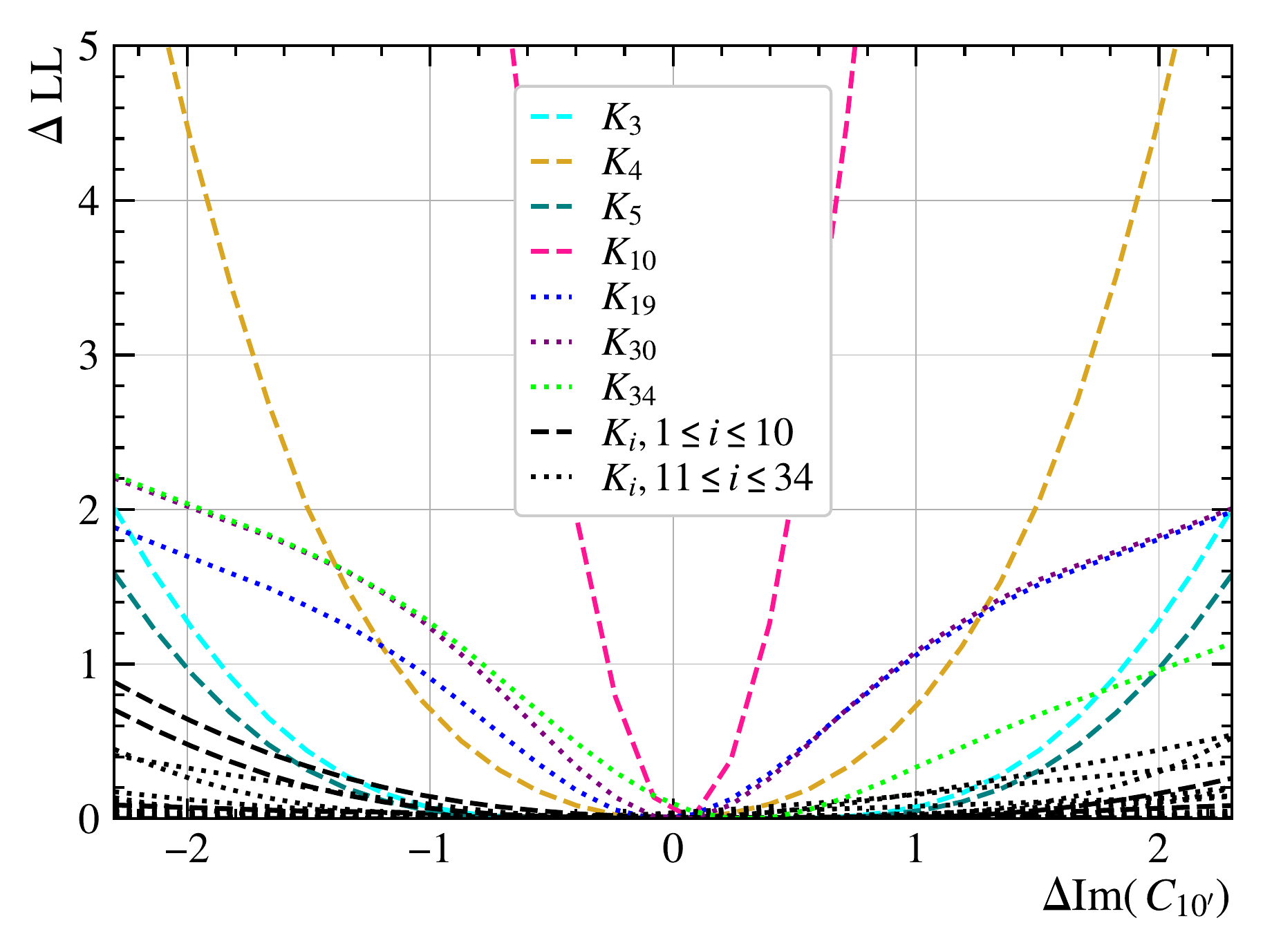}
    \caption{Sensitivity of a fit to the real and imaginary parts of the Wilson coefficients ${C}_{9^\prime}$ and ${C}_{10^\prime}$ to each observable individually. Observables for which the logarithmic likelihood difference is less than one across the shown range are plotted as black lines. Dashed (dotted) lines correspond to observables that are independent of (depend on) the polarization.}
    \label{fig:individual:prime}
\end{figure}

\twocolumngrid

%% file: sensitivity_K_bin1.tex
\begin{tabular}{r|r|c|ccc}
\toprule
$i$ & Value & Stat. & Tot. $\infty\sigma$ & Tot. $1\sigma$ & Tot. $0\sigma$ \\
\midrule
1 & $0.455$ & 0.001 & 0.002 & 0.002 & 0.005 \\
2 & $0.089$ & 0.003 & 0.003 & 0.003 & 0.010 \\
3 & $-0.068$ & 0.002 & 0.003 & 0.004 & 0.007 \\
4 & $-0.332$ & 0.005 & 0.017 & 0.022 & 0.044 \\
5 & $-0.066$ & 0.004 & 0.011 & 0.011 & 0.013 \\
6 & $0.053$ & 0.004 & 0.006 & 0.006 & 0.008 \\
7 & $0.002$ & 0.007 & 0.007 & 0.007 & 0.007 \\
8 & $0.036$ & 0.004 & 0.008 & 0.008 & 0.010 \\
9 & $-0.001$ & 0.007 & 0.007 & 0.007 & 0.007 \\
10 & $-0.000$ & 0.004 & 0.004 & 0.004 & 0.005 \\
11 & $0.163$ & 0.006 & 0.012 & 0.019 & 0.050 \\
12 & $-0.035$ & 0.005 & 0.006 & 0.007 & 0.011 \\
13 & $0.033$ & 0.004 & 0.005 & 0.005 & 0.006 \\
14 & $-0.126$ & 0.010 & 0.014 & 0.015 & 0.020 \\
15 & $0.028$ & 0.007 & 0.012 & 0.012 & 0.011 \\
16 & $-0.024$ & 0.007 & 0.010 & 0.010 & 0.010 \\
17 & $0.018$ & 0.012 & 0.013 & 0.013 & 0.013 \\
18 & $-0.000$ & 0.007 & 0.009 & 0.009 & 0.010 \\
19 & $-0.002$ & 0.012 & 0.012 & 0.012 & 0.012 \\
20 & $-0.001$ & 0.007 & 0.008 & 0.008 & 0.008 \\
21 & $0.002$ & 0.008 & 0.008 & 0.008 & 0.008 \\
22 & $-0.003$ & 0.005 & 0.005 & 0.005 & 0.005 \\
23 & $0.086$ & 0.008 & 0.027 & 0.028 & 0.030 \\
24 & $-0.102$ & 0.005 & 0.005 & 0.006 & 0.011 \\
25 & $-0.003$ & 0.012 & 0.012 & 0.012 & 0.012 \\
26 & $0.003$ & 0.007 & 0.007 & 0.007 & 0.007 \\
27 & $-0.067$ & 0.012 & 0.024 & 0.025 & 0.027 \\
28 & $0.077$ & 0.007 & 0.008 & 0.008 & 0.009 \\
29 & $0.001$ & 0.004 & 0.004 & 0.004 & 0.004 \\
30 & $0.001$ & 0.006 & 0.006 & 0.006 & 0.006 \\
31 & $0.000$ & 0.004 & 0.006 & 0.006 & 0.006 \\
32 & $-0.033$ & 0.006 & 0.033 & 0.032 & 0.025 \\
33 & $-0.001$ & 0.006 & 0.006 & 0.006 & 0.006 \\
34 & $-0.001$ & 0.006 & 0.006 & 0.006 & 0.006 \\
\bottomrule
\end{tabular}

%% file: sensitivity_K_bin2.tex
\begin{tabular}{r|r|c|ccc}
\toprule
$i$ & Value & Stat. & Tot. $\infty\sigma$ & Tot. $1\sigma$ & Tot. $0\sigma$ \\
\midrule
1 & $0.356$ & 0.004 & 0.005 & 0.007 & 0.030 \\
2 & $0.288$ & 0.007 & 0.008 & 0.012 & 0.060 \\
3 & $-0.232$ & 0.006 & 0.006 & 0.012 & 0.042 \\
4 & $-0.237$ & 0.008 & 0.032 & 0.047 & 0.110 \\
5 & $-0.178$ & 0.013 & 0.026 & 0.029 & 0.041 \\
6 & $0.213$ & 0.011 & 0.029 & 0.036 & 0.059 \\
7 & $-0.027$ & 0.014 & 0.018 & 0.018 & 0.022 \\
8 & $-0.107$ & 0.008 & 0.021 & 0.024 & 0.036 \\
9 & $-0.001$ & 0.014 & 0.014 & 0.014 & 0.014 \\
10 & $0.002$ & 0.008 & 0.008 & 0.008 & 0.008 \\
11 & $0.032$ & 0.009 & 0.009 & 0.013 & 0.062 \\
12 & $-0.099$ & 0.014 & 0.015 & 0.015 & 0.014 \\
13 & $0.124$ & 0.012 & 0.012 & 0.012 & 0.019 \\
14 & $-0.021$ & 0.014 & 0.016 & 0.015 & 0.035 \\
15 & $0.087$ & 0.022 & 0.028 & 0.030 & 0.041 \\
16 & $-0.074$ & 0.019 & 0.025 & 0.027 & 0.034 \\
17 & $-0.058$ & 0.023 & 0.026 & 0.027 & 0.030 \\
18 & $-0.005$ & 0.013 & 0.017 & 0.018 & 0.020 \\
19 & $0.000$ & 0.023 & 0.024 & 0.024 & 0.023 \\
20 & $-0.002$ & 0.012 & 0.012 & 0.012 & 0.012 \\
21 & $-0.004$ & 0.015 & 0.015 & 0.015 & 0.015 \\
22 & $0.003$ & 0.008 & 0.008 & 0.008 & 0.009 \\
23 & $0.140$ & 0.014 & 0.016 & 0.015 & 0.014 \\
24 & $-0.153$ & 0.008 & 0.008 & 0.012 & 0.034 \\
25 & $-0.003$ & 0.024 & 0.024 & 0.024 & 0.024 \\
26 & $-0.002$ & 0.013 & 0.013 & 0.013 & 0.013 \\
27 & $-0.120$ & 0.024 & 0.029 & 0.033 & 0.051 \\
28 & $0.103$ & 0.013 & 0.027 & 0.032 & 0.043 \\
29 & $-0.001$ & 0.009 & 0.009 & 0.009 & 0.009 \\
30 & $-0.001$ & 0.010 & 0.010 & 0.010 & 0.010 \\
31 & $0.004$ & 0.009 & 0.011 & 0.011 & 0.015 \\
32 & $0.022$ & 0.010 & 0.027 & 0.023 & 0.016 \\
33 & $0.024$ & 0.009 & 0.009 & 0.010 & 0.013 \\
34 & $-0.004$ & 0.009 & 0.009 & 0.009 & 0.009 \\
\bottomrule
\end{tabular}